\documentclass[fleqn,usenatbib]{mnras}

\usepackage[T1]{fontenc}

\DeclareRobustCommand{\VAN}[3]{#2}
\let\VANthebibliography\thebibliography
\def\thebibliography{\DeclareRobustCommand{\VAN}[3]{##3}\VANthebibliography}

\usepackage{graphicx}	
\usepackage{amsmath}	
\usepackage{amssymb}	
\usepackage{multirow}
\usepackage{newtxtext,newtxmath}

\title[H$\alpha$ Morphologies of Star Clusters in 16 Galaxies]{H$\alpha$ Morphologies of Star Clusters in 16 LEGUS Galaxies: Constraints on HII region evolution timescales}

\author[Hannon S. et al.]{Stephen Hannon,$^{1,2}$\thanks{E-mail: shann004@ucr.edu}
Janice C. Lee,$^{2,3}$
B.C. Whitmore,$^{4}$
B. Mobasher,$^{1}$
\newauthor
D. Thilker,$^{4}$
R. Chandar,$^{5}$
A. Adamo,$^{6}$
A. Wofford,$^{7}$
R. Orozco-Duarte,$^{7}$
D. Calzetti,$^{8}$
\newauthor
L. Della Bruna,$^{6}$
K. Kreckel,$^{9}$
B. Groves,$^{10}$
A. T. Barnes,$^{11}$
M. Boquien,$^{12}$
F. Belfiore,$^{13}$
\newauthor
S. Linden$^{14}$
\\
$^{1}$Department of Physics $\And$ Astronomy, University of California, Riverside, CA, USA\\
$^{2}$Gemini Observatory/NSF’s NOIRLab, 950 N. Cherry Avenue, Tucson, AZ, 85719, USA\\
$^{3}$Steward Observatory, University of Arizona, Tucson, AZ 85721, USA\\
$^{4}$Space Telescope Science Institute, Baltimore, MD, USA\\
$^{5}$Department of Physics and Astronomy, University of Toledo, Toledo, OH, USA\\
$^{6}$Department of Astronomy, The Oskar Klein Centre, Stockholm University, Stockholm, Sweden\\
$^{7}$Instituto de Astronomia, Universidad Nacional Autonoma de Mexico, Unidad Academica en Ensenada, Km 103 Carr. Tijuana-Ensenada, Ensenada, Mexico \\
$^{8}$Department of Astronomy, University of Massachusetts, Amherst, MA 01003, USA\\
$^{9}$Astronomisches Rechen-Institut, Zentrum für Astronomie der Universität Heidelberg, Mönchhofstraße 12-14, 69120 Heidelberg, Germany\\
$^{10}$International Centre for Radio Astronomy Research, University of Western Australia, 7 Fairway, Crawley, 6009, WA, Australia\\
$^{11}$Argelander-Institut f\"{u}r Astronomie, Universit\"{a}t Bonn, Auf dem H\"{u}gel 71, 53121, Bonn, Germany\\
$^{12}$Centro de Astronom\'ia (CITEVA), Universidad de Antofagasta, Avenida Angamos 601, Antofagasta, Chile\\
$^{13}$INAF – Osservatorio Astrofisico di Arcetri, Largo E. Fermi 5, I-50157, Firenze, Italy\\
$^{14}$Astronomy Department, University of Virginia, 530 McCormick Road, Charlottesville, VA 22904, USA\\
}
\date{Accepted XXX. Received YYY; in original form ZZZ}

\pubyear{2021}

\begin{document}
\label{firstpage}
\pagerange{\pageref{firstpage}--\pageref{lastpage}}
\maketitle

\begin{abstract}
The analysis of star cluster ages in tandem with the morphology of their HII regions can provide insight into the processes that clear a cluster’s natal gas, as well as the accuracy of cluster ages and dust reddening derived from Spectral Energy Distribution (SED) fitting. We classify 3757 star clusters in 16 nearby galaxies according to their H$\alpha$ morphology (concentrated, partially exposed, no emission), using Hubble Space Telescope (\textit{HST}) imaging from the Legacy ExtraGalactic Ultraviolet Survey (LEGUS). We find: 1) The mean SED ages of clusters with concentrated (1-2 Myr) and partially exposed HII region morphologies (2-3 Myr) indicate a relatively early onset of gas clearing and a short (1-2 Myr) clearing timescale. 2) The reddening of clusters can be overestimated due to the presence of red supergiants, which is a result of stochastic sampling of the IMF in low mass clusters. 3) The age-reddening degeneracy impacts the results of the SED fitting -- out of 1408 clusters with $M_*\geq$ 5000 M$_{\odot}$, we find that at least 46 (3\%) have SED ages which appear significantly underestimated or overestimated based on H$\alpha$ and their environment, while the total percentage of poor age estimates is expected to be several times larger. 
4) Lastly, we examine the dependence of the morphological classifications on spatial resolution.  At \textit{HST} resolution, our conclusions are robust to the distance range spanned by the sample (3-10 Mpc). However, analysis of ground-based H$\alpha$ images shows that compact and partially exposed morphologies frequently cannot be distinguished from each other.

\end{abstract}




\section{Introduction}
\label{sec:section1}

Clouds of cold gas coalesce to form stars, most of which are thought to form in clustered regions within giant molecular clouds (GMCs; \citealt{LADA03}). Thus an understanding of the star formation cycle in galaxies must encompass a model for the complete evolution of these GMCs, including the formation of stars and star clusters as well as subsequent dispersal of the stars into the ``field'' (e.g., \citealt{WHITMORE14}). One of the key observations regarding GMCs is that the timescale on which these clouds convert all of their gas into stars (i.e. depletion time) has been observed to be much longer than the dynamical time of the cloud. 
In fact only a few percent of GMC mass is converted into stars before the clouds are disrupted \citep{ZUCKERMAN74,WILLIAMS97,EVANS99,KRUMHOLZ07,EVANS09}. 

Despite the long-standing history of these observations, the exact processes that govern these cloud-scale dynamics are still debated, and stellar feedback in general is at the forefront of the discussion (see \citealt{KRUMHOLZ19} for a review). Stellar feedback such as supernovae (SNe) explosions, stellar winds, photoionization, and radiation pressure have been shown to be capable of disrupting the parent molecular cloud, thus halting star formation and adequately limiting star formation efficiency to the observed degree (e.g. \citealt{AGERTZ13,DALE14,HOPKINS18,KRUIJSSEN19}). Further, when simulations exclude stellar feedback, the resulting star formation rate and efficiency are far greater than the aforementioned observations (\citealt{BOURNAND10,TASKER11,HOPKINS11,DOBBS11,KRUMHOLZ11}). 

These feedback mechanisms influence the interstellar medium (ISM) through the transfer of energy and momentum, which contribute to the observed dynamical equilibrium in the ISM \citep{SUN20} among other galactic-scale properties (e.g., \citealt{LEROY08,GOVERNATO10,OSTRIKER10,KRUMHOLZ18}). Importantly, these feedback processes occur on different timescales. Here, we seek to provide constraints on the timescale for gas clearing in order to illuminate the primary feedback mechanism(s) responsible for cloud dispersal, as this process influences the properties of the stellar population (see \citealt{DOBBS14} \& \citealt{ADAMO17} for reviews), ISM, and of galaxy evolution as a whole \citep{SCHAYE15}. 


One of the methods to constrain such timescales is through the study of nebular H$\alpha$ emission in conjunction with the parent star clusters (e.g. \citealt{WHITMORE11,HOLLYHEAD15,HANNON19}). Upon the formation of star clusters within molecular clouds, massive OB-type stars within the clusters will ionize their natal gas cloud and produce HII regions.
A large diversity in the size, shape, and extent of HII regions has been observed via H$\alpha$ imaging (e.g. \citealt{KENNICUTT84,CHURCHWELL06,WHITMORE11,ANDERSON14,HANNON19}), and has been attributed to factors including gas density distributions and the aforementioned stellar feedback mechanisms (\citealt{KRUMHOLZ19}). 
We can then connect the HII morphologies to the parent star clusters' properties such as age, whereby the clusters can effectively be used as clocks to age-date the evolutionary state of the HII regions. Comparison between the morphologies of HII regions and the properties of their parent star clusters can therefore inform us about the HII region evolutionary sequence and its timescale, thus helping us better understand the gas clearing process. 


Previous studies have examined correlations between H$\alpha$ morphology and star cluster properties (e.g. \citealt{WHITMORE11,HOLLYHEAD15,HANNON19}). In particular, in \citet{HANNON19}, we use imaging from the Hubble Space Telescope (\textit{HST}), to study $\sim$700 young ($\leq$ 10 Myr) star clusters in the three nearest (d $\approx$ 4 Mpc) spiral galaxies (NGC 1313, NGC 4395 and NGC 7793) from the Legacy Extragalactic UV Survey (LEGUS; \citealt{CALZETTI15}). 
There we found that star clusters with young ages ($\sim$3 Myr) are associated with concentrated H$\alpha$ emission. 
We confirmed the presence of an evolutionary sequence from concentrated HII regions to more bubble-like regions as reported by \citet{WHITMORE11} and \citet{HOLLYHEAD15}. By comparing the median ages of clusters with these two types of H$\alpha$ morphologies, we also confirmed that the clearing of gas around star clusters begins early (cluster age $\lesssim$ 3 Myr) and takes place over a short interval ($\lesssim$ 1 Myr).


Beyond the study of H$\alpha$ emission, \citet{GRASHA18} and \citet{GRASHA19} compared ALMA (\textit{Atacama Large Millimeter Array}) CO detections of giant molecular clouds (GMCs) with the positions of star clusters in NGC 7793 (one of the galaxies in this study) and found that the timescale for star clusters to dissociate from their natal clouds is similarly short, between 2-4 Myr. This is also roughly consistent with the timescales found based on ALMA CO data for the Antennae galaxies \citep{MATTHEWS18} as well as with \citet{KIM21}, who studied CO(1-0), Spitzer 24$\mu$m, and H$\alpha$ emission in tandem for nearby galaxies. As supernovae (SNe) typically ignite after $\sim$3.5 Myr (e.g. \citealt{AGERTZ13}), these studies support the notion that SNe cannot be the sole driver of gas removal and thus suggest the importance of other clearing mechanisms such as stellar winds, direct radiation pressure, thermal pressure from 10$^4$ K gas, and dust processed radiation pressure (e.g. \citealt{BARNES20} and references therein).

A notable challenge in this type of analysis is that star cluster properties can be affected by stochastic sampling of the upper part of the initial mass function (IMF), following the predictions of previous modeling (e.g. \citealt{BARBARO77, GIRARDI93, LANCON00, BRUZUAL02, CERVINO06, DEVEIKIS08}). These sampling effects greatly affect the luminosity and color distributions of star clusters, which depend strongly on the mass distribution of stars within them (\citealt{FOUESNEAU10,FOUESNEAU12}). For example, \citet{HANNON19} found that $\sim$33$\%$ of their star clusters without H$\alpha$ contained one or more bright red point-like sources, presumably red supergiants, within their photometric apertures. These luminous stars can dominate the flux of lower-mass clusters, resulting in an overestimate of the reddening 
and an underestimate of their ages. Stochastic sampling effects 
have motivated previous star cluster studies (e.g. \citealt{WHITMORE11}; \citealt{HOLLYHEAD15}; \citealt{ADAMO17}) to employ a $\geq$ 5000 M$_{\odot}$ limit, as these more massive clusters more completely sample the IMF and thus 
their SEDs should be less affected by individual bright stars.


One of the major disadvantages of the \citet{HANNON19} sample is the lack of these ``massive" clusters – only 42 clusters were found to have stellar masses above 5000 M$_\odot$, including a mere 12 clusters with concentrated or partially exposed H$\alpha$ morphologies. Further, none of those 12 were isolated from other nearby clusters (defined as clusters with no other clusters within 75 pc), which can produce confounding effects in the presence and clearing of HII regions. To address this, we build upon the sample from \citet{HANNON19}, where the present study quadruples the number of young, massive clusters.


Also, the analysis of \citet{HANNON19} ignores all star clusters with SED ages $>$ 10 Myr under the assumption that H$\alpha$ will only be present when the clusters have ages $<$ 10 Myr (e.g. \citealt{DOPITA06,WHITMORE11,WHITMORE14,HOLLYHEAD15}). 
However, such an assumption ignores the impact of the age-reddening degeneracy, and studies including \citet{HANNON19}, \citet{WHITMORE20}, and \citet{TURNER21} have all identified star clusters for which SED ages and E(B-V) appear to be poorly estimated (e.g. age is overestimated while reddening is underestimated and vice versa). \citet{WHITMORE20} further report their age estimates to have an uncertainty of $\sim$0.3, or a factor of 2 in most cases, but much larger for older clusters such as globular clusters. 
By examining all clusters regardless of SED age, the present study may prove valuable in 
checking the reliability of SED-fit ages, and obtaining new insights regarding timescales associated with gas clearing.


This paper is organized as follows. In Section~\ref{sec:section2} we summarize the observations and the star cluster catalogs used in the study. Section~\ref{sec:section3} reviews the visual classification scheme used to determine H$\alpha$ morphologies, as also employed in \citet{HANNON19}. Section~\ref{sec:section4} examines the SED-estimated cluster age, reddening, and mass distributions as a function of H$\alpha$ morphological class. Sections~\ref{sec:section72} \& \ref{sec:section73} examine manifestations of stochastic sampling and the accuracy of SED-fit ages for two subsets of clusters. The recoverability of star cluster properties as well as resolution dependence on H$\alpha$ morphological classification are examined in Section~\ref{sec:section5}.
Section~\ref{sec:section8} provides an overall summary of the work.

\section{Data}
\label{sec:section2}

\begin{figure}
\includegraphics[width =0.5\textwidth]{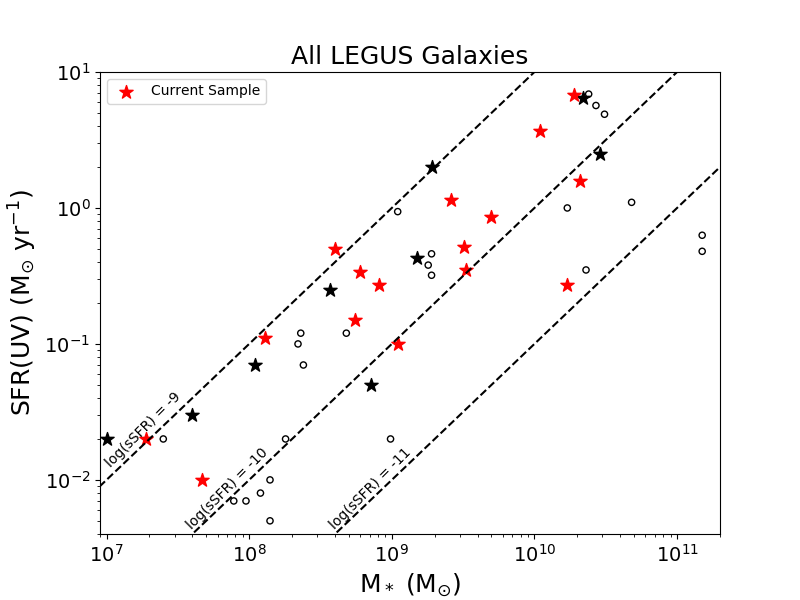}
\caption{Star Formation Rate (SFR) vs. Stellar Mass (M$_*$) for the 16 galaxies studied in this paper (red stars).  The parent sample of 50 LEGUS galaxies \citep{CALZETTI15} are shown while the stars represent the galaxies observed in the LEGUS-H$\alpha$ \textit{HST} follow-up survey, which have relatively higher sSFRs. The 16 red stars show the specific galaxies in this study while the 9 black stars represent galaxies with LEGUS H$\alpha$ imaging but lack available star cluster catalogs. Our sample spans the full range of LEGUS-H$\alpha$ galaxies. The black circles represent the remaining 25 LEGUS galaxies which were not part of LEGUS-H$\alpha$. Dashed lines at log(sSFR) = -9, -10, and -11 are provided for reference.}
\label{fig:SFRvMass}
\end{figure}

The data used in this study are taken from LEGUS \citep{CALZETTI15}, which combines WFC3 (GO-13364; F275W, F336W, F438W, F555W, and F814W) and archival ACS\footnote{To produce consistent final data products, the ACS/WFC images, with a native pixel scale of 0.049$\arcsec$, are aligned to the UVIS WCS reference frame and re-drizzled to the UVIS pixel scale of 0.0396$\arcsec$ \citep{CALZETTI15}.} (F435W, F555W, F606W, and F814W) \textit{HST} imaging to provide full five-band NUV-U-B-V-I coverage for 50 nearby ($\lesssim$ 11 Mpc) galaxies. The LEGUS-H$\alpha$ follow-up survey (GO-13773; PI R. Chandar) provides coverage of the H$\alpha$ emission-line with a narrow-band filter (F657N) as well as a medium-band filter sampling line-free continuum (F547M). This follow-up H$\alpha$ survey was completed for 25 LEGUS galaxies with higher specific star formation rates in the sample (Figure \ref{fig:SFRvMass}). The $\sim$0.04 $\arcsec$/pixel scale of \textit{HST} imaging can resolve objects as small as $\sim$1-3 pc across the distance range of the LEGUS galaxies ($\sim$3--11 Mpc), which allows us to effectively resolve ionizing star clusters and their HII regions, 
a necessary aspect of this analysis.

To produce H$\alpha$ emission-line images,\footnote{Note that the emission-line images also contain flux from the adjacent [NII] 6548,83 lines.
} 
the F657N images are drizzled and aligned, and then an appropriate combination of F814W and F547M, scaled by their AB zeropoints relative to F657N and weighted by their relative pivot wavelengths, are subtracted as a representation of the continuum. 

\begin{figure*}
\includegraphics[width =0.85\textwidth]{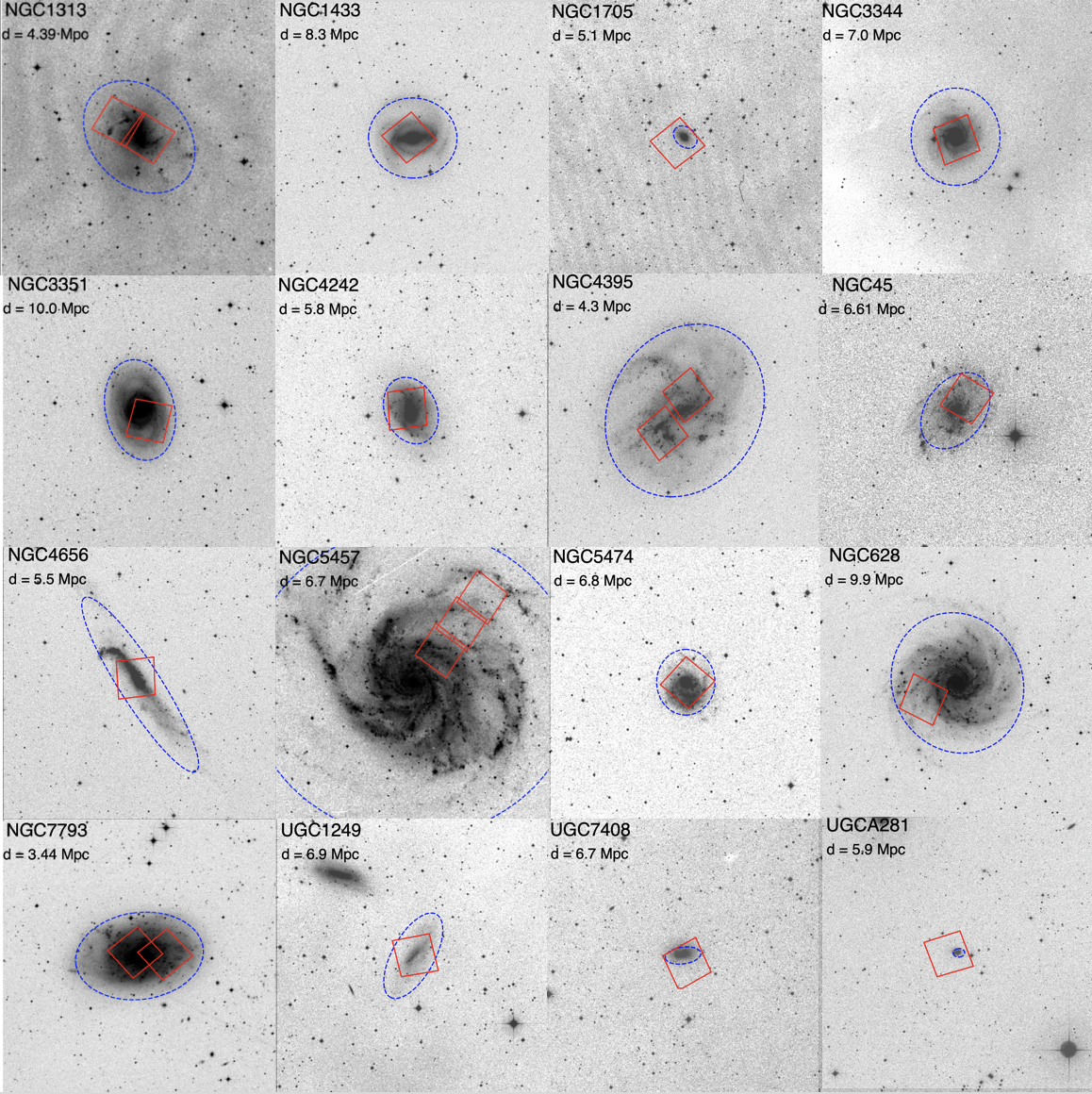}
\caption{\textit{HST} footprints on digitized sky survey (DSS) images for each of the 16 galaxies in this study. Red squares outline the fields of view for F657N (H$\alpha$+[NII] emission-line + continuum) WFC3 images, where galaxies with multiple squares indicate those with multiple fields. Galaxy name and distance are provided in the upper-left of each footprint, and the dashed blue ellipses trace the R$_{25}$ of each galaxy. The field of view of WFC3 is 162$\arcsec$ $\times$ 162$\arcsec$, and each DSS footprint is 20$\arcmin$ $\times$ 20$\arcmin$. Note that LEGUS has aimed to maximize sampling along galactic radii, which is perhaps most evident for NGC 5457.}
\label{fig:Footprints}
\end{figure*}

This study expands upon the work of \citet{HANNON19}, which examined only the three nearest spiral galaxies (6 fields) of the LEGUS sample: NGC 1313, NGC 4395, and NGC 7793. Here we have expanded the sample to include all LEGUS fields which have both cluster catalogs and LEGUS-H$\alpha$ emission-line and continuum imaging. NGC 5457 has also been included, however its cluster catalogs have instead been obtained via private communication and will be published in Linden et al. (in prep.). This has increased the number of galaxies from 3 to 16, and the total number of \textit{HST} fields from 6 to 21. The current dataset also spans a much greater distance range, from $\sim$3--10 Mpc, while the three spiral galaxies examined in \citet{HANNON19} all have distances of $\sim$3--4 Mpc. General properties for each of these galaxies including distance, position angle, inclination, and diameter, are provided in Table~\ref{tab:table1}. 162$\arcsec$ $\times$ 162$\arcsec$ footprints of F657N (imaged with WFC3) are shown on 20$\arcmin$ $\times$ 20$\arcmin$ DSS images for all 16 galaxies in Figure~\ref{fig:Footprints}.

\begin{table*}
\begin{tabular}{lllllllll}
\multicolumn{9}{c}{\textbf{General Properties of Galaxies in the Sample}} \\
\hline
Field   & Type & Distance & Position Angle & Inclination & D$_{25}$ & Z & SFR(UV) & M$_*$  \\
 & & (Mpc) & ($^{\circ}$) & ($^{\circ}$) & (log(')) & & (M$_{\odot}$ yr$^{-1}$) & (M$_{\odot}$) \\
\hline
NGC 1313* & SBd & 4.39           & 40             & 40.7        & 1.04 & 0.02 & 1.15 & 2.6e+09  \\
NGC 1433 & SBab & 8.3            & 88             & 24.5        & 0.27 & 0.02 & 0.27 & 1.7e+10 \\
NGC 1705  & SA0/BCG & 5.1            & 50             & 42.3        & 0.27 & 0.004 & 0.11 & 1.3e+08  \\
NGC 3344  & SABbc & 7.0              & 0              & 23.8        & 0.83 & 0.02 & 0.86 & 5.0e+09 \\
NGC 3351 & SBb & 10.0             & 13             & 47.4        & 0.86 & 0.02 & 1.57 & 2.1e+10 \\
NGC 4242  & SABdm & 5.8            & 25             & 40.4        & 0.58 & 0.02 & 0.10 & 1.1e+09 \\
NGC 4395*  & SAm & 4.3            & 147            & 33.6        & 0.62  & 0.004 & 0.34 & 6.0e+08  \\
NGC 45   & SAdm & 6.61           & 142            & 46.1        & 0.79 & 0.02 & 0.35 & 3.3e+09 \\
NGC 4656  & SBm & 5.5            & 33             & 78.5        & 0.81 & 0.004 & 0.50 & 4.0e+08 \\
NGC 5457  & SABcd & 6.7            & 90             & 20.9        & 1.38  & 0.02, 0.008$^a$ & 6.72 & 1.9e+10  \\
NGC 5474  & SAcd & 6.8            & 0              & 26.5        & 0.38  & 0.02 & 0.27 & 8.1e+08 \\
NGC 628  & SAc & 9.9            & 25             & 25.2        & 1.00 & 0.02 & 3.67 & 1.1e+10 \\
NGC 7793*  & SAd & 3.44           & 98             & 47.4        & 1.02 & 0.02 & 0.52 & 3.2e+09 \\
UGC 1249  & SBm & 6.9            & 150            & 63.4        & 0.81 & 0.02 & 0.15 & 5.5e+08 \\
UGC 7408  & IAm & 6.7            & 100            & 62.5        & 0.30 & 0.004 & 0.01 & 4.7e+07 \\
UGCA 281  & Sm & 5.9            & 95             & 40.7        & -0.11 & 0.004 & 0.02 & 1.9e+07 \\
\hline
\end{tabular}
\caption{General properties for the 16 galaxies examined in this study. This includes RC3 morphological type as listed in HyperLeda (http://leda.univ-lyon1.fr; \citealt{MAKAROV14}), distance, star formation rate, and stellar mass \citep{CALZETTI15}, position and inclination angles \citep{LEE10}, galactic diameter (D$_{25}$; \citealt{MAKAROV14,DEVACOULEURS91}), and metallicity \citep{ADAMO17}. Distances and metallicities match those used to derive physical properties as published in the LEGUS star cluster catalogs. Asterisks indicate the galaxies studied in \citet{HANNON19}. \newline $^a$Z = 0.02 for NGC 5457-NW1, Z = 0.008 for NGC 5457-NW2 and NGC 5457-NW3.}
\label{tab:table1}
\end{table*}

As described in detail in \citet{ADAMO17}, for each of the 21 fields, photometry is performed with aperture radii of 4, 5, or 6 pixels, which correspond to radii of 3.3-7.7 pc, given a pixel scale of 0.03962 $\arcsec$/pixel for WFC3 UVIS. A 1-pixel-wide annulus at a 7-pixel inner radius is then used to subtract the background, with the final cluster magnitudes corrected for foreground Galactic extinction \citep{SCHLAFLY11}.

From the broadband \textit{HST} photometry (i.e., excluding the narrow and medium bands), cluster SEDs are fitted to Yggdrasil SSP models \citep{ZACKRISSON11} with 46 total ages consisting of 1 Myr steps from 1-15 Myr, 10 Myr steps from 20-100 Myr, 100 Myr steps up to 1 Gyr, and lastly 1 Gyr steps up to 14 Gyr. The models also include 150 fixed steps (0.01 mag) in E(B-V) from 0.00 to 1.50 mag.  These SSP models include nebular flux via photoionization modeling with CLOUDY \citep{FERLAND13}, assuming a covering fraction of 0.5 as a fixed parameter \citep{ADAMO17}. 
Stellar libraries assume a fully-sampled \citet{KROUPA01} IMF with stellar masses between 0.1 and 100 M$_{\odot}$, and the fitting for each galaxy adopts the present day metallicity of its young stellar population as determined by nebular abundances \citep{ADAMO17}. 

The cluster ages, E(B-V)s, and masses presented in this work correspond to the minimum $\chi^{2}$ value of the fitting algorithm. Errors for these SED-fitted parameters are determined by the minimum and maximum values found amongst those with $\chi^{2}$ $\leq$ $\chi^{2}_{min}$ + 2.3, based on 1$\sigma$ confidence levels \citep{LAMPTON76}. Detailed descriptions of the SED-fitting algorithm and error analysis are given in \citet{ADAMO10} and \citet{ADAMO12}, respectively. 

A total of twelve cluster catalogs have been made for each of these fields, and each of the catalogs provides a unique set of SED-fit properties. Each catalog assumes one of two stellar evolutionary models (Padova or Geneva; see \citealt{starburst99}, \citealt{VAZQUEZ05}), one of three extinction models (Milky Way; \citealt{CARDELLI89}, Starburst, or Differential Starburst; \citealt{CALZETTI00}), and one of two aperture correction methods \citep[average aperture correction or concentration index-based]{ADAMO17, COOK19}. The rationale for these choices amongst other, more general details are described in \citet{KRUMHOLZ15} \& \citet{ADAMO17}. In this study we examine star cluster properties based on the six models which use standard average aperture correction. It should also be noted that the catalogs using the starburst extinction model were made as a test, however their results are consistent with the other catalogs. A key difference between this study and \citet{HANNON19} is that here we do not narrow the cluster sample to only include clusters with SED-fit ages $\leq$ 10 Myr; we consider all clusters in this analysis, regardless of SED-fit age.

For our study, we use all objects with a visually assigned cluster class of 1, 2, or 3, as determined by the mode of value provided by three classifiers. Cluster classes of 1, 2, and 3 represent symmetric, asymmetric, and multi-peaked star clusters, respectively. We are careful to note, however, that cluster class 3 objects are instead often referred to as compact associations of stars 
\citep{KRUMHOLZ19}. The term 'clusters' will be inclusive of these objects throughout this paper, except for explicitly stated comparisons. The cluster classes are provided in the LEGUS cluster catalogs (\citealt{ADAMO17, COOK19}; Linden et al. in prep.) and reveal an even distribution of cluster class 1 (31.5\%), 2 (35.9\%), and 3 (32.6\%) objects for the sample. For a more in-depth analysis on the distribution of these cluster morphologies in LEGUS galaxies, see \citet{GRASHA15,GRASHA17}. 



Overall, we present results based on 3757 final cluster targets across 16 galaxies totaling 21 fields for which we have H$\alpha$ emission-line data. This includes $\sim$1900 clusters with SED ages $\leq$ 10 Myr and $\sim$170 clusters with SED ages $\leq$ 10 Myr and stellar masses $\geq$ 5000 M$_\odot$. The total number of objects as well as the number of young and young+massive clusters for each of the fields in this study are shown in Table~\ref{tab:table2}.

\begin{table*}
\begin{tabular}{llllll}
\multicolumn{6}{c}{\textbf{Detection \& Cluster Counts by Field}} \\
\hline
\textbf{Field}      & \textbf{Total Detections} & \textbf{Verified Clusters} & \textbf{$\leq$ 10Myr} & \textbf{$\geq$5000 $M_{\odot}$} & \textbf{$\geq$10000 $M_{\odot}$} \\
\hline
NGC1313E*   & 5284             & 259               & 115             & 15                    & 9                      \\
NGC1313W*   & 3784             & 486               & 180             & 19                    & 13                     \\
NGC1433    & 1099             & 168               & 123             & 14                    & 8                      \\
NGC1705    & 365              & 42                & 10              & 2                     & 1                      \\
NGC3344    & 1527             & 396               & 270             & 8                     & 1                      \\
NGC3351    & 1389             & 292               & 169             & 42                    & 33                     \\
NGC4242    & 1080             & 191               & 83              & 1                     & 0                      \\
NGC4395N*   & 291              & 39                & 25              & 0                     & 0                      \\
NGC4395S*   & 837              & 137               & 115             & 7                     & 4                      \\
NGC45      & 774              & 117               & 58              & 0                     & 0                      \\
NGC4656    & 1703             & 262               & 122             & 20                    & 9                      \\
NGC5457NW1 & 5891             & 283               & 92              & 5                     & 3                      \\
NGC5457NW2 & 3707             & 110               & 40              & 2                     & 2                      \\
NGC5457NW3 & 5390             & 36                & 18              & 3                     & 1                      \\
NGC5474    & 4275             & 177               & 89              & 2                     & 0                      \\
NGC628E    & 593              & 259               & 120             & 21                    & 6                      \\
NGC7793E*   & 899              & 191               & 117             & 2                     & 1                      \\
NGC7793W*   & 2794             & 221               & 132             & 1                     & 0                      \\
UGC1249    & 1000             & 88                & 51              & 3                     & 1                      \\
UGC7408    & 305              & 46                & 13              & 0                     & 0                      \\
UGCA281    & 327              & 15                & 9               & 2                     & 2  \\
\hline
\textbf{TOTAL}      &   \textbf{43314}               & \textbf{3815}              & \textbf{1951}            & \textbf{169}                   & \textbf{94}     \\    
\hline
\end{tabular}
\caption{Number counts of total detections, visually-verified clusters (LEGUS mode class = 1, 2, or 3), verified clusters with SED age $\leq$10Myr, and verified clusters with SED age of $\leq$10Myr \& stellar mass of either $\geq$5000M$_{\odot}$ or $\geq$10000M$_{\odot}$ for each of the 21 fields of view. The six fields which were studied in \citet{HANNON19} are denoted with an asterisk. NGC 5457 catalogs have been obtained via private communication and are to be published by Linden et al. (in prep).}
\label{tab:table2}
\end{table*}

\section{H$\alpha$ Morphology Classification}
\label{sec:section3}

\begin{figure}
\includegraphics[width =0.47\textwidth]{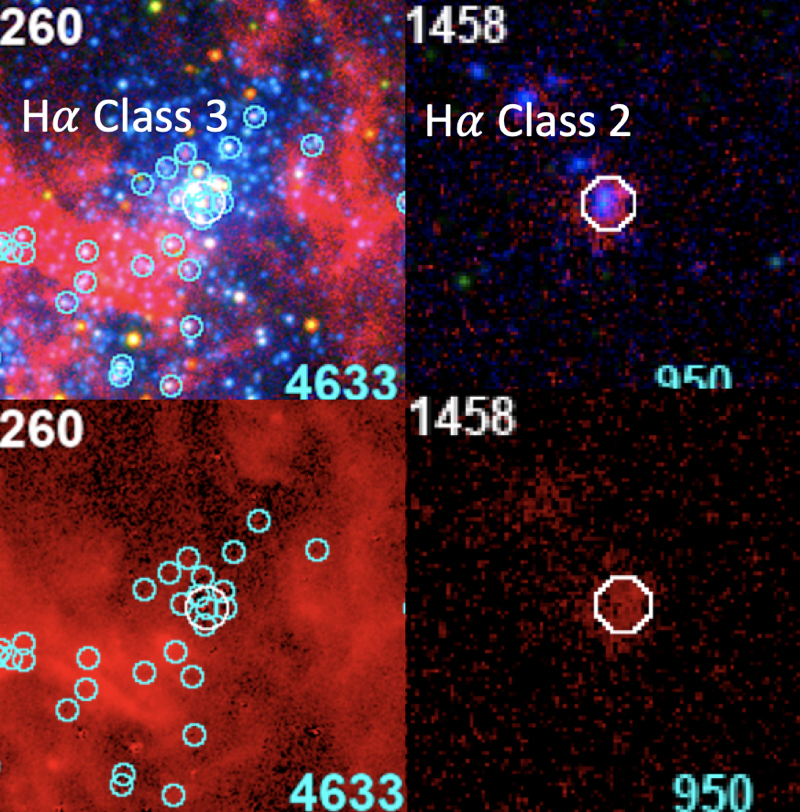}
\caption{150 pc $\times$ 150 pc postage stamps highlighting two of the more difficult H$\alpha$ morphology classifications, with the final classes indicated in the image. The top row is made up of the RGB images of the clusters while the corresponding continuum-subtracted H$\alpha$ is shown in the bottom row. Object IDs and cluster mass (in M$_{\odot}$) are provided in the upper-left and lower-right of each stamp, respectively. Crowded star cluster regions can produce widely dispersed and irregular H$\alpha$ emission which can make the distinction between an H$\alpha$ class of 2 versus 3 difficult for clusters such as Object 260. Faint emission can also make classifications harder to determine, such as Object 1458 which appears to show a small amount of clearing on the left side of the cluster.}
\label{fig:ClassDifficulty}
\end{figure}

Following the visual classification process of \citet{HANNON19}, we create two sets of 150 pc $\times$ 150 pc images centered on each cluster, which we refer to as "postage stamps". One set of postage stamps shows only the continuum-subtracted H$\alpha$ in red, which allows us to most clearly determine the shape and extent of each HII region. The other set of stamps is made from a composite RGB image of the galaxy using combined NUV and U bands, combined V and I bands, and the continuum-subtracted H$\alpha$ narrow band for the blue, green, and red channels, respectively. These allow us to examine the HII regions in the context of the target cluster and any other neighboring objects. Examples of these postage stamps can be found throughout this paper, firstly in Figure~\ref{fig:ClassDifficulty}.


The definition of each of the three H$\alpha$ morphologies is as follows, as introduced in \citet{WHITMORE11}:

\renewcommand{\labelenumi}{\arabic{enumi}}
\begin{enumerate}
    \item \textbf{Concentrated (H$\alpha$ Class 1)}, where the target star cluster has H$\alpha$ emission covering it, and where there are no discernible bubbles or areas around the cluster which lack H$\alpha$ emission,
    \item \textbf{Partially exposed (H$\alpha$ Class 2)}, where the H$\alpha$ surrounding the target cluster displays bubble like or filamentary morphology covering part of the cluster and,
    \item \textbf{No emission (H$\alpha$ Class 3)}, where the target cluster appears to be clear of H$\alpha$, i.e. there is no H$\alpha$ emission within $\sim$20 pc of the cluster. This includes those which have no visible H$\alpha$ emission in their entire 150 pc-wide postage stamps.
\end{enumerate}

Clear examples of each of these H$\alpha$ morphologies can be found in Figure 2 of \citet{HANNON19}. 

For reference, the 5$\sigma$ point source detection limit for the H$\alpha$ images used in this study is between 5.0x10$^{-17}$ and 5.5x10$^{-17}$ ergs cm$^{-2}$ s$^{-1}$. Across our distance range ($\sim$3--10 Mpc), the detectable observed luminosities are thus $\sim$7.8x10$^{34}$ -- 6.6x10$^{35}$ erg s$^{-1}$. Given the model grid of \citet{SMITH02B}, these luminosities correspond to the ionizing fluxes of B0.5V and B0V stars for the nearest and furthest galaxies, respectively. These calculations assume solar metallicity, Case B recombination, nebular temperatures and densities of 10$^4$ K and 100 cm$^{-3}$, respectively, and that the nebular region is radiation bounded.

During the initial classification of each postage stamp, the shape and position of H$\alpha$ relative to the target cluster is taken into consideration according to the above definitions of each of the classes. To facilitate this classification process, collages of postage stamps for all clusters in a given field, ordered by best-fit cluster age, are first examined. For $\sim$80$\%$ of clusters, the classification of H$\alpha$ morphology is relatively unambiguous, such as when a cluster resides in a dense pocket of H$\alpha$ emission (concentrated morphology) or when there is hardly any H$\alpha$ emission in its entire postage stamp (no emission). However, the remaining $\sim$20$\%$ are not so obvious, such as when the H$\alpha$ emission is faint or when the target cluster is found in a crowded region where the distribution of H$\alpha$ emission is uneven. An example of the latter case would be Object 260 (given an H$\alpha$ class of no emission; Figure~\ref{fig:ClassDifficulty}), where there is plenty of H$\alpha$ emission in the vicinity due to a large number of clusters, but the H$\alpha$ is distributed quite irregularly and not in a circular pocket or bubble. This uneven distribution makes it more difficult to determine the origin and extent of the H$\alpha$ emission, and thus whether it should be identified as an H$\alpha$ class of 2 or 3. Object 1458 (Figure~\ref{fig:ClassDifficulty}) shows what appears to be a cluster that is partially exposed from the H$\alpha$ in the RGB image, however upon inspection of the H$\alpha$-only image, one could argue that there are no obvious cleared regions within the H$\alpha$ emission and should be classified as an H$\alpha$ class 1 instead. Thus the faint H$\alpha$ emission of Object 1458 makes its classification difficult. 

After these initial classifications are made, a second inspection is performed by the same person (SH). For each of the 21 fields, postage stamps are put into separate collages according to the three initial H$\alpha$ classes. In this manner, a more difficult case is able to be directly compared to the "obvious" cases in that particular H$\alpha$ class in order to discern whether the former in fact belongs in the class. Clear misidentifications are very occasionally made on the initial classifications and are subsequently corrected on this second pass. These corrections are performed until the H$\alpha$ morphologies in the collages for each of the classes are satisfactorily uniform. For the clusters in NGC 1313, NGC 4395, and NGC 7793, H$\alpha$ morphologies from \citet{HANNON19} are treated as initial classifications and are thus subjected to this second inspection as well.

By virtue of having the same person go through the classification process twice, an error rate can be drawn on the classification accuracy of the inspector themself. As shown in Table~\ref{tab:table3}, 48 (1.3$\%$) of the clusters in this sample had their H$\alpha$ morphology classes changed upon second inspection, 6 of which were obvious mistakes (0.2$\%$). The majority (83$\%$) of these changes occur for clusters that are not isolated, which is defined here as having at least one other star cluster within 75 pc.\footnote{This separation is based on \citet{GOULIERMIS18}, who  studied several thousand unbound stellar systems in Local Group galaxies and found that the average size of these stellar associations is $\sim$70-90pc.} Our figures indicate that, while the number of H$\alpha$ class changes that occur is quite small, the crowded regions naturally present the most difficulty. We have verified that these uncertainties do not affect the results discussed in following sections.

\begin{table}
\centering
\begin{tabular}{lll}
\multicolumn{3}{c}{\textbf{H$\alpha$ Morphology Reclassifications}} \\
\hline
H$\alpha$ Class Change & Isolated & Not Isolated \\
\hline
1 to 2                 & 2        & 6            \\
1 to 3                 & 0        & 2            \\
2 to 1                 & 4        & 13           \\
2 to 3                 & 1        & 0            \\
3 to 1                 & 0        & 1            \\
3 to 2                 & 1        & 18   
\\
\hline
Total                  & 8        & 40
\\
\hline
\end{tabular}
\caption{A total of 48 clusters had their initial H$\alpha$ morphological class changed upon second inspection. The class changes for these objects are presented here, divided by isolation criteria and the specific class change. The 6 possible changes are shown in the first column (initial classification to new classification), where 1, 2, and 3 represent concentrated, partially exposed, and no emission H$\alpha$ morphologies, respectively.}
\label{tab:table3}
\end{table}





Table~\ref{tab:table4} lists the final overall distribution of H$\alpha$ morphological classes amongst the 3757 clusters in the sample. In total, there are 499 clusters with concentrated H$\alpha$ (180 isolated, 319 non-isolated), 372 clusters with partially exposed H$\alpha$ (108 isolated, 264 non-isolated), and 2886 clusters with no H$\alpha$ emission (1757 isolated, 1129 non-isolated). Looking at those that have best-fit ages $\leq$ 10 Myr and comparing them to the equally-defined cluster sample of \citet{HANNON19}, we find that the number of clusters with concentrated H$\alpha$, partially exposed H$\alpha$, and no emission have all roughly tripled, from 142 to 478 (3.4x), from 112 to 327 (2.9x), and to 400 to 1112 (2.8x), respectively. Thus the relative ratios of clusters in each H$\alpha$ morphological class have remained consistent across these two studies. Further, the ratio of isolated to non-isolated clusters within each H$\alpha$ morphological class has also remained roughly consistent: 36, 13, and 47 percent of clusters with concentrated H$\alpha$, partially exposed H$\alpha$, and no emission, respectively, were found to be isolated in \citet{HANNON19} compared to 36, 30, and 54 percent of clusters in the present study. The most notable change is the relative increase in isolated clusters with partially exposed H$\alpha$, however this could be attributed to the smaller sample size of \citet{HANNON19}, which contained only 14 such clusters. 

The complete list of H$\alpha$ classifications is provided as supplementary online material, and an abbreviated version of this table can be found in Appendix A.

\begin{table*}
\begin{tabular}{lllllll}
\multicolumn{7}{c}{\textbf{H$\alpha$ Morphology Classifications by Field}} \\
\hline
                      & \multicolumn{2}{c}{\textbf{Concentrated}} & \multicolumn{2}{c}{\textbf{Partially Exposed}} & \multicolumn{2}{c}{\textbf{No Emission}} \\
\hline
Field      & Isolated & Not Isolated & Isolated & Not Isolated & Isolated & Not Isolated \\
\hline
NGC1313E   & 8          & 19             & 2          & 28             & 94         & 101            \\
NGC1313W   & 3          & 29             & 0          & 26             & 118        & 305            \\
NGC1433    & 13         & 19             & 13         & 26             & 69         & 28             \\
NGC1705    & 1          & 2              & 1          & 5              & 14         & 19             \\
NGC3344    & 37         & 60             & 17         & 38             & 136        & 97             \\
NGC3351    & 16         & 34             & 18         & 18             & 153        & 53             \\
NGC4242    & 6          & 12             & 4          & 6              & 126        & 37             \\
NGC4395N   & 6          & 0              & 1          & 0              & 29         & 3              \\
NGC4395S   & 4          & 27             & 3          & 30             & 32         & 41             \\
NGC45      & 9          & 5              & 3          & 2              & 90         & 8              \\
NGC4656    & 10         & 20             & 7          & 12             & 97         & 116            \\
NGC5457NW1 & 14         & 7              & 16         & 9              & 171        & 65             \\
NGC5457NW2 & 8          & 3              & 0          & 4              & 91         & 3              \\
NGC5457NW3 & 2          & 0              & 0          & 0              & 30         & 2              \\
NGC5474    & 12         & 12             & 6          & 5              & 115        & 26             \\
NGC628E    & 9          & 20             & 5          & 11             & 165        & 37             \\
NGC7793E   & 9          & 17             & 3          & 14             & 74         & 66             \\
NGC7793W   & 5          & 23             & 4          & 23             & 71         & 85             \\
UGC1249    & 6          & 5              & 4          & 5              & 44         & 24             \\
UGC7408    & 0          & 0              & 0          & 0              & 33         & 13             \\
UGCA281    & 2          & 5              & 1          & 2              & 5          & 0              \\
\hline
\textbf{Total}      & \textbf{180}        & \textbf{319}            & \textbf{108}        & \textbf{264}            & \textbf{1757}       & \textbf{1129}           \\
\hline
\textbf{$\leq$10Myr}      & 171        & 307            & 99         & 228            & 601        & 511            \\
\textbf{$\geq$5000M$_{\odot}$}     & 4          & 35             & 8          & 21             & 42         & 53             \\
\textbf{$\geq$10000M$_{\odot}$}    & 1          & 26             & 5          & 13             & 22         & 26 \\
\hline
\end{tabular}
\caption{Number counts of clusters in each of the three H$\alpha$ morphology classes (concentrated, partially exposed, no emission) and further separated into isolated and non-isolated categories depending on the presence or absence of neighboring clusters within 75pc. Also included at the bottom of the table are the total number of clusters in each morphology \& isolation bin that are $\leq$10Myr,  $\leq$10Myr \& $\geq$5000M$_{\odot}$, and $\leq$10Myr \& $\geq$10000M$_{\odot}$. These numbers do not include 58 clusters which were found outside of the field of view for F657N.}
\label{tab:table4}
\end{table*}

\section{Age, E(B-V), and Mass Distributions}
\label{sec:section4}

To directly compare the results of this study to those of \citet{HANNON19}, we first limit the cluster sample to those which have best-fit SED ages $\leq$ 10 Myr as before. 
We then investigate the results without any age restrictions to examine the impact of including clusters with older age estimates, which may include young clusters with erroneous ages (e.g. \citealt{WHITMORE20}).

For each of the six LEGUS catalogs of SED-fit properties, we examine the statistics of cluster age, reddening, and mass, according to H$\alpha$ morphology, and whether the cluster is isolated. Our results are presented in the following subsections. 




\subsection{Age Statistics}
\label{sec:section41}

\begin{figure*}
\includegraphics[width =0.85\textwidth]{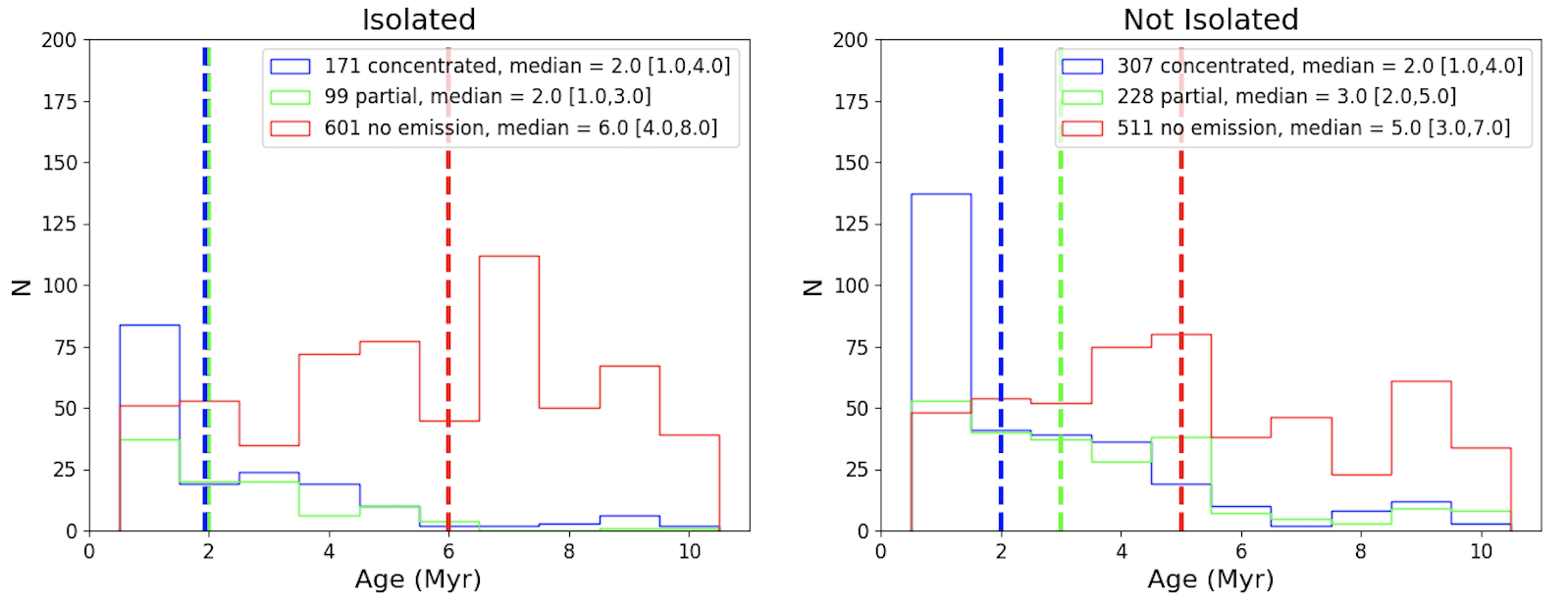}
\caption{Age histograms for young ($\leq$ 10 Myr) clusters with concentrated H$\alpha$ (blue), partially exposed H$\alpha$ (green), and no emission (red). The left and right plots show the distributions for isolated and non-isolated clusters, respectively. Vertical dashed lines show the median SED cluster age for each of the morphological classes, where the blue and green vertical lines overlap for isolated clusters due to the equal median ages of those with concentrated and partially exposed H$\alpha$ morphologies. 
While we find similarity in the age distributions of clusters with concentrated and partially exposed H$\alpha$, those with concentrated H$\alpha$ have an excess in the lowest age bin (1 Myr). As expected, those without H$\alpha$ have age distributions shifted to older ages.}
\label{fig:AgeHist}
\end{figure*}

Analysis of the SED age distributions for each of the H$\alpha$ morphology classes is pertinent to understanding whether or not these classes constitute an evolutionary sequence. Figure~\ref{fig:AgeHist} shows the age distributions for each of the H$\alpha$ morphological classes for clusters with SED ages $\leq$ 10 Myr based on our reference sample (i.e., SED fits assuming Padova stellar evolutionary model, Milky Way extinction). 


For the reference sample of isolated clusters with SED ages $\leq$ 10 Myr, we find that the median best-fit age [first quartile, third quartile] of the 171 clusters with concentrated HII regions is 2.0 [1.0, 4.0] Myr, while the median ages for the 99 partially exposed clusters and 601 clusters with no H$\alpha$ emission are 2.0 [1.0, 3.0] Myr and 6.0 [4.0, 8.0] Myr, respectively. If we include all clusters regardless of best-fit age, we find that the median SED ages of the clusters with concentrated and partially exposed H$\alpha$ are unchanged (2.0 [1.0, 4.0] Myr, 2.0 [1.0, 4.0] Myr, respectively), while the median SED age of clusters without H$\alpha$ is significantly older (50.0 [7.0, 200.0] Myr), as would be expected. Additionally, we find relatively consistent results for the sample of non-isolated clusters.


Thus we find that the inclusion of clusters with SED ages $>$ 10 Myr does not significantly affect the median ages of clusters with concentrated and partially exposed H$\alpha$, largely due to the fact that there are relatively few of these older clusters. 21 of the 499 clusters (4.2$\%$) with concentrated H$\alpha$ and 45 of the 372 clusters (12.1$\%$) with partially exposed H$\alpha$ have best-fit ages $>$ 10 Myr. Although few, these clusters are of considerable interest as these ages are beyond the expected lifetime of ionizing stars, and are discussed in detail in Section~\ref{sec:OMCs}. 



Across all stellar evolutionary models (Padova, Geneva) and extinction/attenuation models (Milky Way, Starburst, Differential Starburst), and whether or not we differentiate between isolated and non-isolated clusters, we find a consistent progression for young clusters: those with concentrated H$\alpha$ have a median SED age of 1-2 Myr, those with partially exposed H$\alpha$ have a median SED age of 2-3 Myr, and those without H$\alpha$ have a median SED age of 3-6 Myr. We also note that the inclusion of 'compact associations' in our sample (Section~\ref{sec:section2}) does not affect this progression. For an analysis on the robustness of the SED-fit ages, we guide the reader to Section~\ref{sec:cigale}.

If we take the error in a cluster's SED age to be the difference between its maximum and minimum age (see Section~\ref{sec:section2}), we find that the median age error for each H$\alpha$ morphological class is 1 Myr. As this is smaller than each of the distributions' standard deviations, we determine that the errors are insufficient in describing the observed distribution widths, thus indicating that the distribution widths likely represent real variations in the cluster ages within each H$\alpha$ morphological class.

While the median SED ages reveal an age progression, we note that there is overlap in the age distributions, particularly for clusters with concentrated and partially exposed H$\alpha$ classes. To mathematically determine the uniqueness of the age distributions for all H$\alpha$ morphological classes, we perform Kolmogorov-Smirnov (KS) tests which calculate the probability that each individual distribution originates from the same parent distribution, with the results summarized in Table~\ref{tab:table7}. KS tests are performed for isolated and non-isolated cluster samples separately, as well as for a combined sample of isolated + non-isolated clusters to improve the overall number statistics.



\begin{table*}
\begin{tabular}{lllllll}
\multicolumn{7}{c}{\textbf{Model Comparison for Star Clusters $\leq$ 10Myr}} \\
\hline
Environment       & \multicolumn{2}{c}{Isolated} & \multicolumn{2}{c}{Non-Isolated} & \multicolumn{2}{c}{All} \\
\hline
                & Padova   & Geneva   & Padova       & Geneva   & Padova   & Geneva   \\
\hline
H$\alpha$ Class & \multicolumn{6}{c}{Age}                                             \\
\hline
1 vs. 2         & 0.322    & 0.019    & <$p_{3\sigma}$     & <$p_{3\sigma}$ & <$p_{3\sigma}$ & <$p_{5\sigma}$ \\
1 vs. 3         & <$p_{5\sigma}$    & <$p_{5\sigma}$    & <$p_{5\sigma}$        & <$p_{5\sigma}$    & <$p_{5\sigma}$    & <$p_{5\sigma}$    \\
2 vs. 3         & <$p_{5\sigma}$    & <$p_{5\sigma}$    & <$p_{5\sigma}$        & <$p_{5\sigma}$    & <$p_{5\sigma}$    & <$p_{5\sigma}$    \\
\hline
                & \multicolumn{6}{c}{E(B-V)}                                          \\
\hline
1 vs. 2         & <$p_{3\sigma}$ & <$p_{3\sigma}$ & <$p_{5\sigma}$        & <$p_{5\sigma}$    & <$p_{5\sigma}$    & <$p_{5\sigma}$    \\
1 vs. 3         & <$p_{3\sigma}$ & <$p_{3\sigma}$ & 9.74E-03         & <$p_{3\sigma}$ & <$p_{3\sigma}$    & $p_{3\sigma}$    \\
2 vs. 3         & <$p_{5\sigma}$    & <$p_{5\sigma}$    & <$p_{3\sigma}$     & <$p_{3\sigma}$ & <$p_{5\sigma}$    & <$p_{5\sigma}$    \\
\hline
                & \multicolumn{6}{c}{Mass}                                            \\
\hline
1 vs. 2         & 0.417    & 0.411    & 0.747        & 0.417     & 0.968    & 0.707    \\
1 vs. 3         & 4.92E-02     & 0.136    & 0.549        & 0.182    & 0.634    & 1.12E-02    \\
2 vs. 3         & 0.642    & 0.274    & 0.871        & 0.920     & 0.889    & 0.280    \\
\hline
\end{tabular}
\caption{p-values from KS tests to compare the H$\alpha$ morphological classes of clusters with SED ages $\leq$10Myr based on their distributions of cluster age (top), reddening (middle), and mass (bottom). These are split based on environment (Isolated, Not Isolated, and All clusters (i.e. regardless of isolation)) as well as whether the Padova or Geneva stellar evolutionary model was used, each assuming Milky Way extinction. ``$p_{3\sigma}$" and ``$p_{5\sigma}$" represent p-values below 2.70E-03 and 5.73E-07, respectively, indicating that the null hypothesis that the two samples are drawn from the same distribution can be rejected at a >3$\sigma$ or >5$\sigma$ confidence level according to the Gaussian error function.}
\label{tab:table7}
\end{table*}

Overall, when the cluster population is divided into isolated and non-isolated samples, KS tests reveal that we cannot reject the hypothesis that the age distributions for clusters with partially exposed H$\alpha$ morphologies have been drawn from the same parent sample as clusters with concentrated H$\alpha$. The age distribution of clusters with no H$\alpha$ emission, however, are found to be statistically different ($\geq$ 5$\sigma$ confidence) from those of the concentrated and partially exposed H$\alpha$ classes, as expected. When the isolated and non-isolated samples are combined to increase the sample sizes, the statistical differences between the age distributions of the concentrated and partially exposed H$\alpha$ classes versus the distribution for clusters without H$\alpha$ remain at the $\geq$ 5$\sigma$ confidence level, while the p-values improve to the $\sim$4$\sigma$ confidence level between the age distributions for clusters with concentrated and partially exposed H$\alpha$ morphologies, similar to the confidence levels for non-isolated clusters. These results are also consistent across all stellar evolutionary models and extinction models. 

The similarity in both the age distributions and median ages of star clusters with concentrated and partially exposed H$\alpha$ morphologies indicate that the gas clearing timescale is $\lesssim$ 1 Myr, in agreement with \citet{HANNON19}. Additionally, we can also determine a clearing timescale based on mean ages. Clusters with concentrated H$\alpha$ have a mean age of 2.7 and 2.3 Myr according to the Padova and Geneva models, respectively, while the clusters with partially exposed H$\alpha$ have a mean age of 3.2 and 2.8 Myr, respectively. If we subtract the ages of each H$\alpha$ class in quadrature (assuming simple gaussian statistics are a reasonable approximation for uncertainties such as differences in physical clearing times, measurement uncertainties, and number statistics), we find an age difference between clusters with concentrated and partially exposed H$\alpha$ of 1.7 and 1.6 Myr based on the Padova and Geneva models, respectively. This provides a rough estimate of the clearing time based on our measurements. Splitting the sample into isolated \& non-isolated groups reveals that most of this difference in mean age comes from non-isolated clusters, for which the same calculation yields age differences between the two H$\alpha$ classes of 2.1 and 1.9 Myr for the Padova and Geneva models, respectively. These figures therefore suggest a clearing timescale of 1-2 Myr rather than $\lesssim$ 1 Myr.

Further, we note that the 10 Myr cutoff is simply a construction of the initial analysis used to limit the number of clusters which require visual inspection for classification (based on the assumption that no H$\alpha$ emission would be found for older clusters), and more significant differences in the age distributions are apparent at the youngest ages (e.g., the peak at 1 Myr for clusters with concentrated H$\alpha$, see Figure~\ref{fig:AgeHist}).
If we instead consider only clusters with SED ages $\leq$ 5 Myr, which marks the 80th percentile in age for clusters with H$\alpha$ (either concentrated or partially exposed), we find that the median SED ages of these clusters become more distinguished: 1.0 [1.0, 3.0] Myr for clusters with concentrated H$\alpha$, and 2.0 [1.0, 4.0] Myr for clusters with partially exposed H$\alpha$.
We also see this reflected in KS tests; the p-values comparing the age distributions for class 1 and 2 H$\alpha$ morphologies 
are reduced by a factor of $\sim$2, although this remains below the 2-sigma confidence level for the isolated sample of clusters. 
With these considerations, we report more generally that the gas clearing timescale is 1-2 Myr, in agreement with other recent works such as \citet{CHEVANCE20}.

As done in \citet{HANNON19}, we can alternatively infer the lifetimes of the H$\alpha$ classes by examining their relative fractions, with the assumption that our sample is statistically representative of all clusters $\leq$ 10 Myr. For these objects with SED ages $\leq$ 10 Myr, the fraction of clusters with class 1, 2 and 3 H$\alpha$ morphologies is 20-30\%, 10-20\%, and 50-70\%, respectively, where the ranges reflect the use of isolated or non-isolated clusters. These fractions imply a lifetime of 2-3 Myr for concentrated H$\alpha$ morphologies, which indicates when the clusters begin to clear their natal gas, and also a lifetime of 1-2 Myr for partially exposed morphologies, which corresponds to the duration of the gas clearing process. Both of these findings are consistent with the results from the median SED ages of each H$\alpha$ class.

It is important to note, however, that our sample is likely not complete amongst clusters embedded within their natal gas clouds (akin to a concentrated H$\alpha$ morphology) as they can be obscured beyond detection (e.g. \citealt{MESSA21}, \citealt{KIM21}), though a more detailed analysis would be necessary than is available here and is thus addressed in future opportunities (Section~\ref{sec:section8}).

\subsection{Reddening Statistics}
\label{sec:section42}

\begin{figure*}
\includegraphics[width =0.85\textwidth]{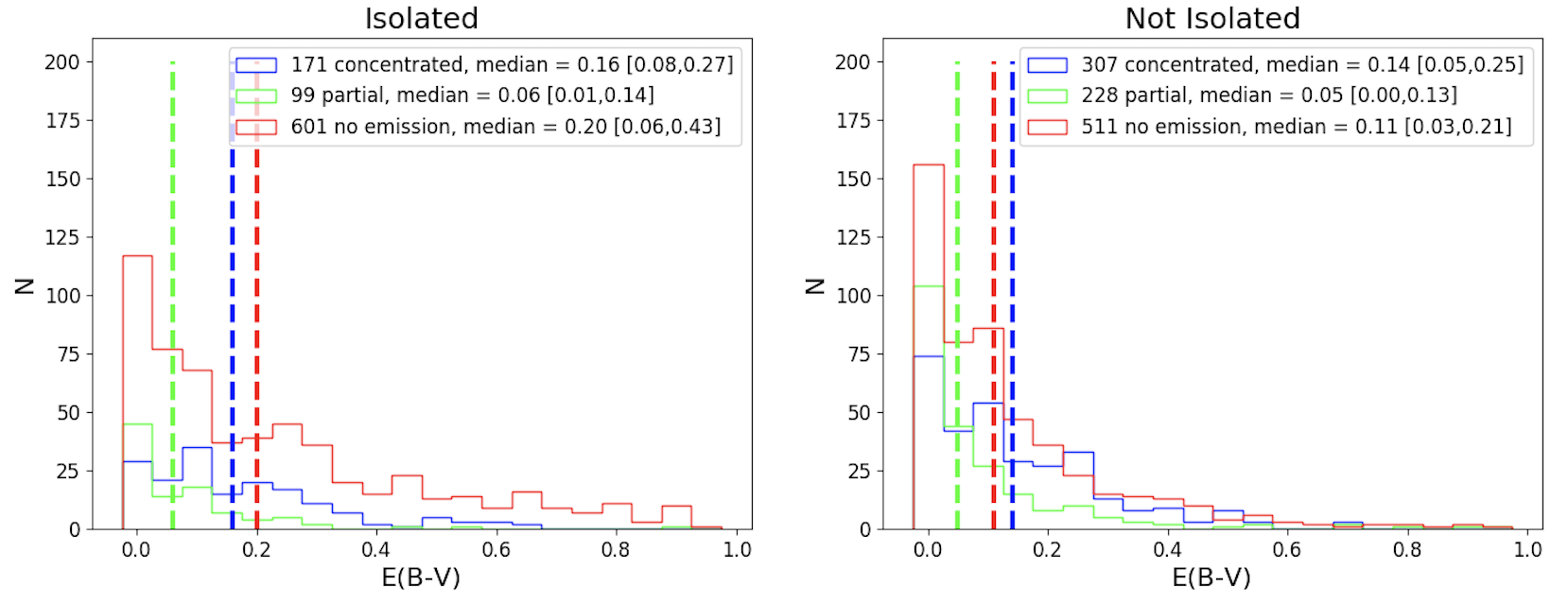}
\caption{Reddening (E(B-V)) histograms for clusters with SED ages $\leq$ 10 Myr and with concentrated H$\alpha$ (blue), partially exposed H$\alpha$ (green), and no emission (red). The left and right plots show the distributions for isolated and non-isolated clusters, respectively. Vertical dashed lines show the median cluster reddening for each of the morphological classes. Comparable reddening distributions are unexpectedly found amongst clusters with concentrated and absent H$\alpha$.}
\label{fig:RedHist}
\end{figure*}

We now examine the reddening of star clusters in each H$\alpha$ morphological class. E(B-V) measures the extinction of a star cluster due to dust, and considering cluster environment and H$\alpha$ morphology, can help to inform us as to the accuracy of the SED-fitting. Figure~\ref{fig:RedHist} shows distributions of E(B-V) values, used to measure the presence of dust, for each of the H$\alpha$ morphological classes for clusters with best-fit ages $\leq$ 10 Myr. These values are derived from five-band SED fitting, assuming Milky Way extinction \citep{CARDELLI89}. 


Contrary to the expectation that H$\alpha$ traces dust content, we find clusters with concentrated H$\alpha$ and those without H$\alpha$ to have comparable reddening estimates, while those with partially exposed H$\alpha$ show the least reddening. For the reference (Padova, Milky Way extinction) sample of isolated clusters with SED ages $\leq$ 10 Myr, the median E(B-V) [first quartile, third quartile] of the 171 clusters with concentrated H$\alpha$ is 0.16 [0.08, 0.27] mag, while the E(B-V) for the 99 partially exposed clusters and 601 clusters with no H$\alpha$ emission are 0.06 [0.01, 0.14] and 0.20 [0.06, 0.43] mag, respectively. If we include all clusters regardless of SED-fit age, we find similar results: E(B-V)s of 0.15 [0.07, 0.26], 0.06 [0.01, 0.15], 0.14 [0.04, 0.33] mag are found for clusters with concentrated, partially exposed, and absent H$\alpha$, respectively. The biggest difference in median E(B-V) here is again between clusters without H$\alpha$, as the inclusion of older clusters triples their sample size, however their median E(B-V) remains larger than that of clusters with partially exposed H$\alpha$ and similar to that of clusters with concentrated H$\alpha$. As also observed in Section 4.1, the sample of non-isolated clusters reveal qualitatively consistent results.


Across all stellar evolutionary models (Padova, Geneva) and extinction models (Milky Way, Starburst, Differential Starburst), and whether or not we delineate between isolated and non-isolated clusters, we find consistent results: clusters with partially exposed H$\alpha$ have the smallest median E(B-V) (0.05-0.08 mag), while those with concentrated H$\alpha$ (0.13-0.16 mag) or no emission (0.11-0.22 mag) have higher, comparable cluster reddening values. These results are also consistent regardless of whether we include compact associations in the cluster sample. 

If we take the error in E(B-V) to be the difference between the maximum and minimum E(B-V) value for an individual cluster (see Section~\ref{sec:section2}), we find that the errors are greater for clusters without H$\alpha$ (0.09-0.10 mag) than for clusters with concentrated (0.06 mag) and partially exposed (0.05) H$\alpha$ morphologies. These greater errors indicate that the relatively large E(B-V) for clusters without H$\alpha$ may not be physical, as discussed in Section~\ref{sec:section72}.

Upon examination of the KS-test results of cluster reddening (Table~\ref{tab:table7}), the distributions of E(B-V) for clusters with partially exposed H$\alpha$ and no emission appear statistically different ($\gtrsim$5$\sigma$) regardless of whether the cluster is isolated. Clusters with concentrated and partially exposed H$\alpha$ morphologies also appear statistically unique at a $>$ 3$\sigma$ confidence level for all models. The E(B-V) distributions for clusters with concentrated H$\alpha$ and no emission, however, are a bit different: When the sample is divided into isolated and non-isolated samples, we find that while some models show $\sim$3$\sigma$ confidence that they are statistically unique, others show significantly lower confidence ($\lesssim$ 1$\sigma$) for both divisions. When we combine the isolated and non-isolated samples, KS tests show that all distributions are significantly different ($>$ 5$\sigma$) except for clusters with concentrated H$\alpha$ and no emission ($\gtrsim$ 3$\sigma$). The comparable reddening estimates found between these two H$\alpha$ classes are examined further in Section~\ref{sec:section72}.

\subsection{Mass Statistics}
\label{sec:section43}
Lastly, we examine the stellar mass of our star clusters. More massive clusters will more completely sample the IMF and hence be more likely to include massive OB stars. These stars are necessary in this study because they ionize their surrounding gas, which produces the H$\alpha$ emission we use to classify HII region morphologies. 



For clusters with best-fit SED ages $\leq$ 10 Myr, the median cluster mass is $\sim$1000 M$_{\odot}$, regardless of the H$\alpha$ classification and whether or not the cluster is isolated. For star clusters of this mass, we are likely to observe effects due to stochastic sampling of the IMF as the colors of the clusters may be disproportionately represented by individual bright stars (e.g. \citealt{FOUESNEAU12}), which in turn can affect the estimated age and E(B-V) determined by the SED-fitting algorithm (e.g. \citealt{HANNON19}), which assumes a fully populated IMF. We investigate these effects further in Section~\ref{sec:section72}.

When we include clusters with SED ages $>$ 10 Myr, we find that the median mass of clusters with concentrated and partially exposed H$\alpha$ remain unchanged ($\sim$1000 M$_{\odot}$), while clusters without H$\alpha$ see their median mass increase to $\sim$4000 M$_{\odot}$. This is likely a selection effect, however, as older objects become less luminous as they age, and thus the minimum detectable cluster mass increases with age for a given magnitude limit (e.g. \citealt{ADAMO17}). A potential complicating factor here is that amongst these more massive objects are likely to be old, globular clusters, as indicated by the significant fraction ($\sim$50\%) of clusters without H$\alpha$ which are symmetric (cluster class = 1). The inclusion of globular clusters is notable because SED-fitting has shown particular difficulty in accurately estimating their ages and reddening (\citealt{TURNER21,WHITMORE20}), and with sufficiently underestimated ages, the globular clusters can contaminate our sample of truly young clusters. Examples of such age underestimations are studied further in Section~\ref{sec:section73}.

\section{H$\alpha$ morphologies and UBVI Color-Color Diagram}
\label{sec:section72}

\begin{figure*}
\includegraphics[width =1.0\textwidth]{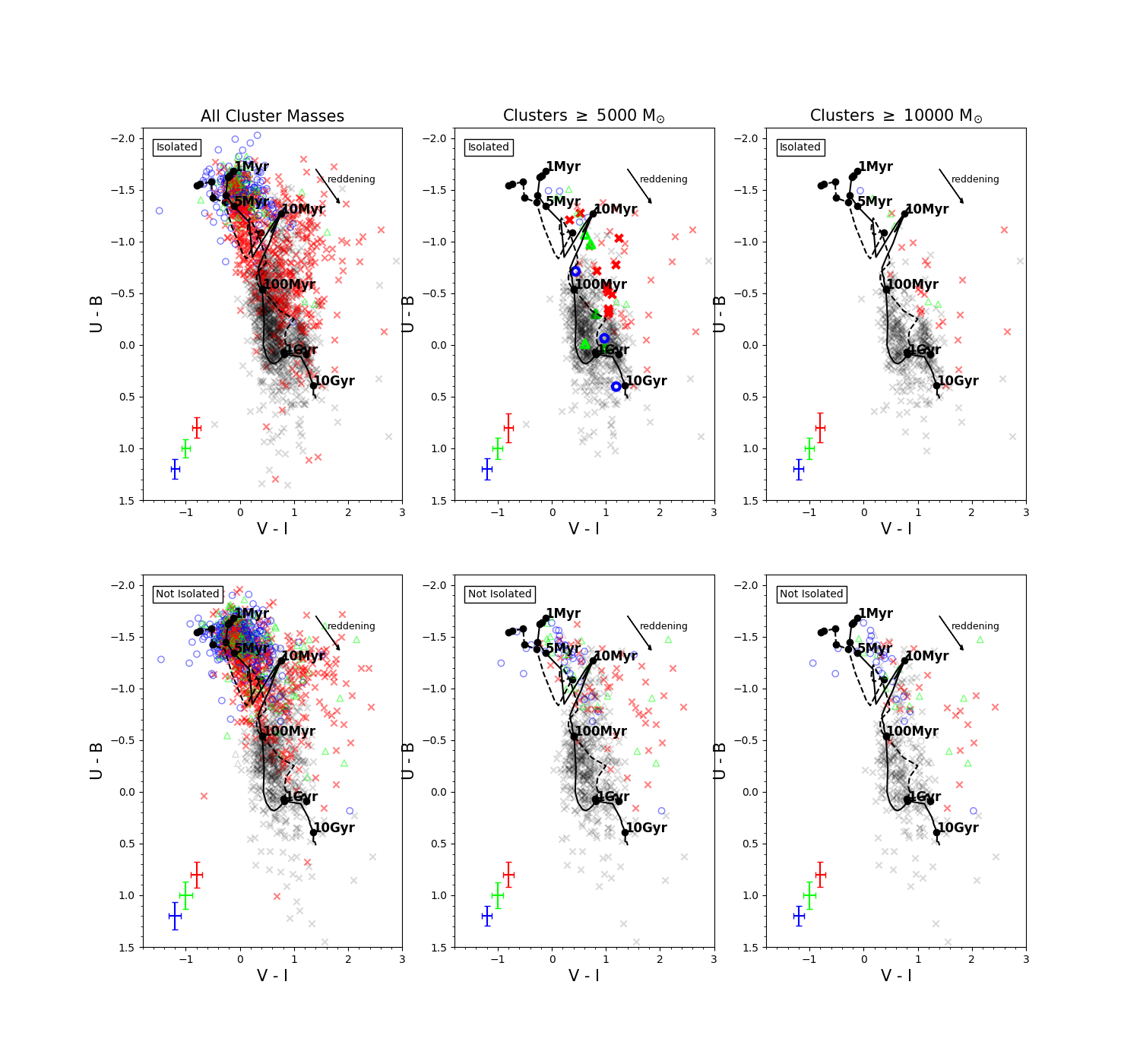}
\caption{(U-B) vs. (V-I) plots of star clusters across all 16 galaxies of the sample. Yggdrasil model tracks (solid line for Z=0.02; dashed line for Z=0.004) used for the fitting of these clusters are included (Table~\ref{tab:table1} lists the metallicity used for each galaxy). 1 Myr, 5 Myr, 10 Myr, 100 Myr, 1 Gyr, and 10 Gyr points along each track are indicated by solid black circles. Blue circles and green triangles, which represent clusters with SED ages ($\leq$ 10 Myr) and concentrated or partially exposed H$\alpha$ morphologies, respectively, are both largely focused near the youngest end of the models. Red Xs, representing clusters with SED ages ($\leq$ 10 Myr) without H$\alpha$, are much more widely dispersed along older sections of the model, as expected. Gray circles, triangles, and crosses represent the same H$\alpha$ morpholgies, but have best-fit ages $>$ 10 Myr. The top and bottom rows of plots show clusters that are isolated and not isolated, respectively. The columns show clusters with no mass limit (left), clusters above 5000 M$_{\odot}$ (middle), and clusters above 10000 M$_{\odot}$ (right). Median photometric errors for each H$\alpha$ morphological class are indicated by crosses in the lower-left of each plot. Bold, colored points in the upper middle plot indicate clusters for which the SED-fit ages seem to be inconsistent with the presence or absence of H$\alpha$; i.e., clusters with SED ages $\leq$ 3 Myr yet no H$\alpha$, and those with associated H$\alpha$ \& SED ages $>$ 10 Myr, as further discussed in Section~\ref{sec:section73}.}
\label{fig:CCs}
\end{figure*}

Examining the colors of individual clusters can provide useful insight into the SED-fitting procedure as well as highlight potential anomalies to further investigate. In Figure \ref{fig:CCs}, we present color-color diagrams ((U-B) vs. (V-I)) for the full sample of clusters, classified by H$\alpha$ morphology, where the colored points represent clusters with SED ages $\leq$ 10 Myr and gray points show clusters with older SED ages ($>$ 10 Myr). In these plots, we note that clusters with concentrated (blue circles) and partially exposed (green triangles) H$\alpha$ morphologies, which have median SED ages of 1-3 Myr, overlap near the youngest end of the stellar evolutionary models. Clusters without H$\alpha$ emission, on the other hand, are found generally toward the older parts of the SSP model track, as expected. 


As in \citep{HANNON19}, we also observe manifestations of stochastic sampling of the IMF. Most notably, the positions of our clusters in UBVI space are consistent with those predicted by the stochastic modeling of low-mass clusters by \citet{FOUESNEAU10}, \citet{FOUESNEAU12}, and also \citet{OROZCO21}. Figure 2 of \citet{FOUESNEAU12} provides an example of these predicted positions in the same color space as Figure~\ref{fig:CCs}, highlighted by a spray of points to the lower-right of the 10 Myr point of the model SSP track. 
Despite all of the colored points in Figure \ref{fig:CCs} representing clusters with SED-fit ages $\leq$ 10 Myr, some are found as far along the SSP model as around its 10 Gyr point, and are interspersed amongst the clusters with SED ages $>$ 10 Myr (gray points). 

In the upper-right corner of each plot, we provide a reddening vector of A$_V$ = 1.0 \citep{CARDELLI89} which, in principle, indicates that the more a cluster is reddened, the further the cluster will be moved from the model, along the direction of that vector (down and to the right). Typically, this reddening is due to dust covering parts or all of the cluster (see Figure 5 of \citealt{HOLLYHEAD15}), however we find in our sample that the clusters with the least evidence of dust (i.e. lacking H$\alpha$ emission) are in fact the ones which appear the most reddened. These occupy the same region as the aforementioned stochastically sampled clusters, and in fact, the color-color plots in Figure 9 of \citet{HANNON19} and Figure 9 of \citet{WHITMORE20} show that this phenomenon is indeed at least partly attributable to stochastic sampling effects. In those cases, the fluxes of clusters in this region are dominated by the presence of bright red point-like sources, presumably red supergiants. In this study, at least one bright red point-like source was found within the photometric aperture of $\sim$24$\%$ of all clusters with SED ages $<$ 10 Myr and without H$\alpha$. This fraction ranges between $\sim$20-40\% with no dependence on galaxy distance. Of all clusters found to the lower-right of the 10 Myr point on the Yggdrasil model (V-I $>$ 0.8, U-B $>$ -1.3), $\sim$70$\%$ 
contain a bright red point-source. 

In turn, these clusters containing bright red sources are found to have a median E(B-V) $\sim$2.5x larger than the clusters without a bright red star (0.26 vs. 0.11 mag, respectively). This indicates that the SED algorithm is likely interpreting the red color from bright stars within a cluster as reddening due to dust, which results in an overestimate of E(B-V) while underestimating its age as a consequence of the age-extinction degeneracy. As noted in Section~\ref{sec:section42}, we see $\sim$2x greater errors in E(B-V) for clusters without H$\alpha$ compared to those with H$\alpha$ (see also \citealt{WHITMORE20} \& \citealt{TURNER21}), and we further find that, of the clusters without H$\alpha$, those containing red point-like sources have a median reduced $\chi^2$ value 3x greater than clusters without red sources. These errors indicate that the stochastic sampling of bright red stars results in poorer fitting of a cluster's SED. It is also noteworthy that these effects are seen even when limiting the sample to clusters $\geq$ 5000 M$_{\odot}$ (middle column of Figure \ref{fig:CCs}) and $\geq$ 10000 M$_{\odot}$ (right column), although the sample sizes are small. Beyond the age-extinction degeneracy, the poor fitting we observe may also be due to the assumption of a fixed metallicity for each galaxy, as older clusters are metal-poor relative to the younger population on which the SED-fitting is based (also see \citealt{DEGER21}).

Another point which was not addressed in \citealt{HANNON19} is that we find clusters blueward of the SSP models (V-I $\lesssim$ -0.3), approximately half of which belong to galaxies with sub-solar assumed metallicities. These clusters are of particular interest because they should be able to be traced backward along the reddening vector (up and to the left) to the SSP model. However, these clusters cannot be traced back in such a fashion as they are already upward and/or leftward of the model. Even more interesting is that nearly all of these clusters have concentrated H$\alpha$, and would thus likely be reddened. We suggest two possible causes for the very blue color of these clusters:

1) As the model tracks assume a gas covering fraction of 0.5, there could be additional contributions from nebular emission if the covering fraction is higher, which would push the color of these clusters blueward, and would especially affect those with sub-solar metallicities. This effect is demonstrated in Figure 1 of \citet{ZACKRISSON01}, where model predictions of V-I and U-B colors are found to be as much as 0.5 mag less and 0.5 mag greater, respectively, for young clusters which use a covering fraction of 1.0 versus 0.5.

2) Stochastic effects may result in the undersampling of low-mass stars which results in the flux-dominance of massive ionizing blue stars.
To this end, \citet{OROZCO21} find in their models that the presence of two Wolf-Rayet stars (types WC or WNE) within low-mass clusters (10$^3$ M$_{\odot}$) can produce this effect.  Notably, all of the clusters from our sample which are in this region have masses $\leq$ 5000 M$_{\odot}$. 

\section{Accuracy of SED-fit Ages}
\label{sec:section73}

Because key conclusions from this study are based on small differences (1 Myr) in median ages, one of the questions we must ask is how accurate the SED ages are. To investigate this accuracy for individual clusters, we look into two particular cases where the data do not align with expectation: 1) very young (SED age $\leq$ 3 Myr) clusters without H$\alpha$ and 2) older (SED age $>$ 10 Myr) clusters with H$\alpha$. The number of these cases in each field are listed in Table \ref{tab:YMCs+OMCs}, where the totals and percentages of their relevant population in the sample are also shown. 

\subsection{\textbf{"Young" Massive Clusters without H$\alpha$}}
\label{sec:YMCs}

We first examine isolated, massive ($\geq$ 5000 M$_{\odot}$) clusters that do not have H$\alpha$ emission, yet have very young ($\leq$ 3 Myr) SED-fit ages. These are interesting because we expect clusters of these ages and masses to have pre-supernova, massive ionizing stars, and thus to have HII regions. We specify isolated clusters in order to avoid confounding effects neighboring clusters could have on the clearing of a cluster's natal gas and hence the morphological class of H$\alpha$. As shown in Table \ref{tab:YMCs+OMCs}, there are only 12 such cases across all 21 fields, however this age restriction means that this represents a lower limit to the overall number of clusters with younger-than-expected SED ages (e.g. \citealt{WHITMORE20,TURNER21}).

\begin{table}
\begin{tabular}{lllll}
           & \multicolumn{2}{l}{Clusters $\leq$3Myr without H$\alpha$} & \multicolumn{2}{l}{Clusters \textgreater{}10Myr with H$\alpha$} \\
\hline
Field      & Isolated                    & Not Isolated                & Isolated                      & Not Isolated                    \\
\hline
NGC1313E   & 0                           & \textbf{\textcolor{red}{3}}                           & \textbf{\textcolor{red}{1}}                             & \textbf{\textcolor{red}{2}}                               \\
NGC1313W   & 0                           & 0                           & 0                             & \textbf{\textcolor{red}{4}}                               \\
NGC1433    & \textbf{\textcolor{red}{8}}                           & 0                           & 0                             & \textbf{\textcolor{red}{4}}                               \\
NGC1705    & 0                           & 0                           & \textbf{\textcolor{red}{1}}                             & \textbf{\textcolor{red}{4}}                               \\
NGC3344    & 0                           & 0                           & 0                             & \textbf{\textcolor{red}{2}}                               \\
NGC3351    & \textbf{\textcolor{red}{2}}                           & \textbf{\textcolor{red}{4}}                           & \textbf{\textcolor{red}{4}}                             & \textbf{\textcolor{red}{5}}                               \\
NGC4242    & 0                           & 0                           & 0                             & 0                               \\
NGC4395N   & 0                           & 0                           & 0                             & 0                               \\
NGC4395S   & 0                           & \textbf{\textcolor{red}{1}}                           & 0                             & \textbf{\textcolor{red}{1}}                               \\
NGC45      & 0                           & 0                           & 0                             & 0                               \\
NGC4656    & 0                           & \textbf{\textcolor{red}{2}}                           & \textbf{\textcolor{red}{1}}                             & 0                               \\
NGC5457NW1 & 0                           & 0                           & \textbf{\textcolor{red}{1}}                             & \textbf{\textcolor{red}{1}}                               \\
NGC5457NW2 & 0                           & \textbf{\textcolor{red}{1}}                           & 0                             & 0                               \\
NGC5457NW3 & 0                           & 0                           & 0                             & 0                               \\
NGC5474    & 0                           & 0                           & 0                             & 0                               \\
NGC628E    & \textbf{\textcolor{red}{1}}                           & \textbf{\textcolor{red}{2}}                           & 0                             & 0                               \\
NGC7793E   & 0                           & 0                           & 0                             & \textbf{\textcolor{red}{1}}                               \\
NGC7793W   & 0                           & 0                           & 0                             & \textbf{\textcolor{red}{1}}                               \\
UGC1249    & \textbf{\textcolor{red}{1}}                           & 0                           & 0                             & 0                               \\
UGC7408    & 0                           & 0                           & 0                             & 0                               \\
UGCA281    & 0                           & 0                           & 0                             & \textbf{\textcolor{red}{1}}                               \\
\hline
\textbf{TOTALS}      & \textbf{12 (63$\%$)}                 & \textbf{13 (26$\%$)}                 & \textbf{8 (40$\%$)}                    & \textbf{26 (32$\%$)}   \\
\hline
\end{tabular}
\caption{Number counts of massive ($\geq$5000 M$_{\odot}$) clusters with best-fit ages $\leq$ 3 Myr without H$\alpha$,  and massive clusters with best-fit ages > 10 Myr with H$\alpha$ (concentrated or partially exposed H$\alpha$ morphology). These are organized by field and whether they are isolated, with their totals shown at the bottom. 63$\%$ and 26$\%$ of massive, young ($\leq$3Myr; before the onset of supernovae) isolated and non-isolated clusters, respectively, are classified as no emission. 40$\%$ and 32$\%$ of massive, isolated and non-isolated clusters with H$\alpha$, respectively, have best-fit ages > 10 Myr, beyond the expected limit for H$\alpha$ association.}
\label{tab:YMCs+OMCs}
\end{table}

\begin{figure*}
\includegraphics[width =1.0\textwidth]{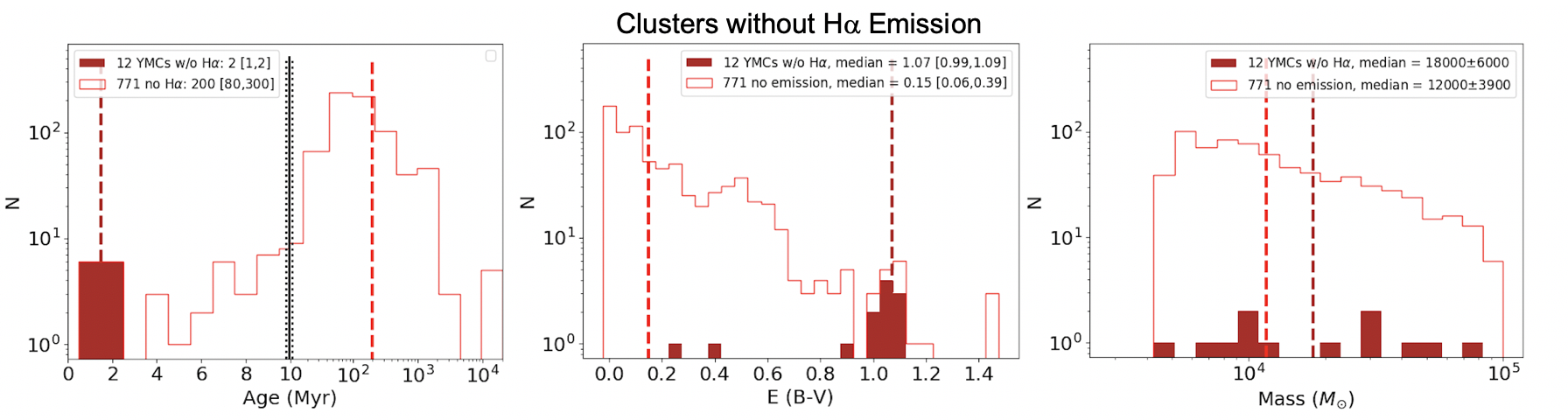}
\caption{Age, E(B-V), and mass histograms of isolated, massive ($\geq$5000 M$_{\odot}$) clusters without H$\alpha$. The light red outline represents the distribution of all of these clusters while the filled, dark red distribution represents our objects of interest: the youngest (SED age $\leq$ 3 Myr), massive clusters without H$\alpha$. The dashed, vertical light red and dark red lines represent the medians of all and young massive clusters without H$\alpha$, respectively. While there is not a strong distinction in the stellar mass of these clusters, the clusters with SED ages $\leq$ 3 Myr are found to have relatively high E(B-V). Note in the leftmost plot that the scale on the x-axis transitions from linear to logarithmic at 10 Myr.}
\label{fig:YMCs_Hist}
\end{figure*}

While these clusters represent the youngest massive clusters without H$\alpha$, we also find that their estimated E(B-V)s are much greater than the larger population. Figure~\ref{fig:YMCs_Hist} shows the age, E(B-V), and mass distributions of all massive, isolated clusters without H$\alpha$, where the filled distributions indicate the 12 with SED ages $\leq$ 3 Myr. The reddening distribution (middle panel) shows that these clusters have a median E(B-V) over 7 times that of the greater sample (1.07 vs. 0.15 mag). The stellar masses of these clusters, however, show a relatively flat distribution across the observed mass range, indicating that they do not have relatively lower masses, which would have suggested a greater likelihood of lacking OB stars and hence H$\alpha$ emission. 

\begin{figure*}
\includegraphics[width =0.85\textwidth]{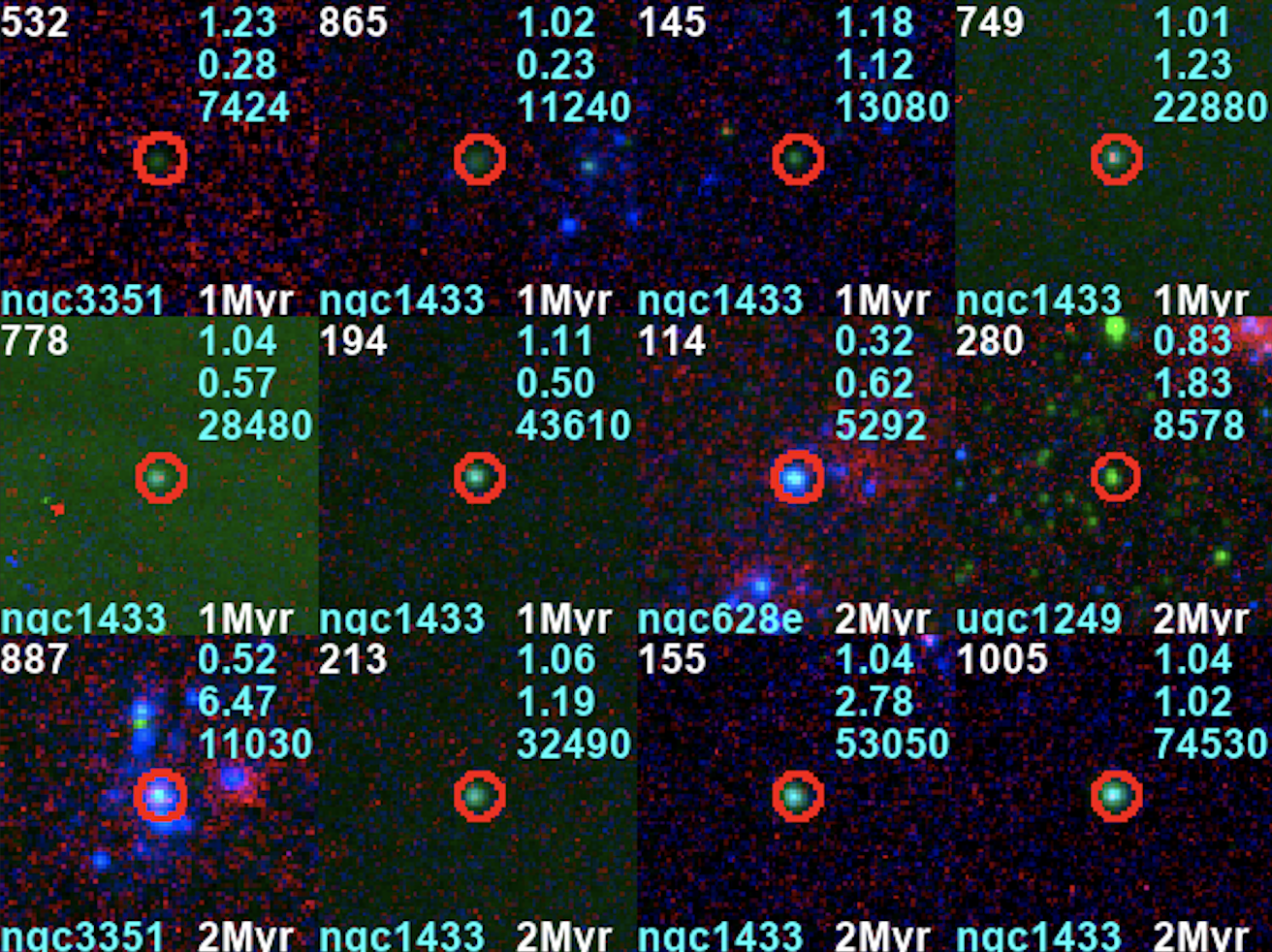}
\caption{150 pc $\times$ 150 pc RGB postage stamps for each of the 12 massive, isolated clusters classified as no emission. Object ID is in the upper-left, field in the lower-left, and SED age in the lower-right of each stamp. In the upper-right corner of each stamp are the E(B-V), $\chi^2$ fit, and mass, from top to bottom.}
\label{fig:YMCs_RGB}
\end{figure*}

By visually inspecting each of these clusters, we can determine how trustworthy their age \& E(B-V) estimates are. Figure~\ref{fig:YMCs_RGB} displays the 150 pc $\times$ 150 pc RGB postage stamps for each of the 12 clusters. Here we note that these 12 objects of interest are spread across 4 galaxies (NGC 628E, UGC 1249, NGC 1433, and NGC 3351), with two-thirds of them coming from NGC 1433. In two of the postage stamps (objects 887 of NGC 3351 and 114 of NGC 628E), we find evidence in support of their young SED ages: 1) their blue (F275W + F336W) color, 2) their proximity to other blue stars, and 3) H$\alpha$ emission is in the vicinity (within the 75 pc radius of the stamp). 
For these clusters, it is possible that they have already cleared their natal gas and/or more proximal H$\alpha$ emission is too faint to detect.

\begin{figure}
\includegraphics[width =0.45\textwidth]{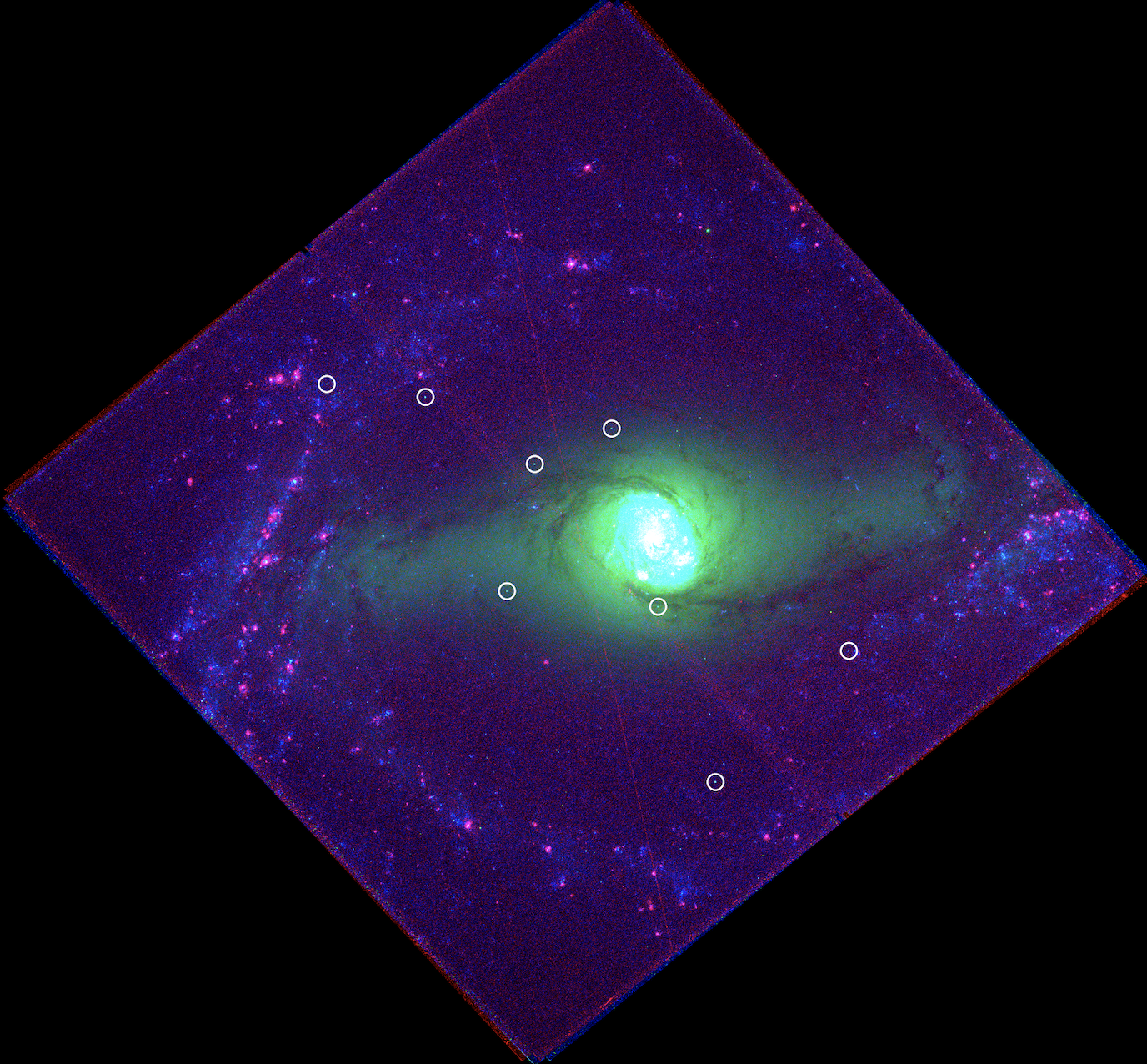}
\caption{Field view of NGC 1433. The locations of clusters without H$\alpha$ but with SED ages $\leq$ 2 Myr and masses $\geq$ 5000 M$_{\odot}$ are denoted by white circles. Proximity to the galaxy's bulge indicates enhanced likelihood that clusters are likely old, globular clusters rather than young clusters.}
\label{fig:ngc1433}
\end{figure}

The remaining 10 clusters, however, have colors consistent with older star clusters and do not display any of the above characteristics. That is, they show little to no blue emission themselves, lack neighboring blue stars (with the exception of object 865 of NGC 1433, which only has a couple), and do not show any nearby H$\alpha$ emission. Additionally, 9 of these objects are classified as symmetric clusters (cluster class = 1), consistent with old, globular clusters. The clusters found in NGC 1433 are displayed in Figure~\ref{fig:ngc1433}, which highlights that many are proximal to the galaxy's bulge, where globular clusters are typically found. Despite the consistencies between these objects and older clusters, we find that the errors in their SED ages are limited: the SED-fitting algorithm has determined maximum ages of $\leq$ 8 Myr for all of the globular cluster candidates, 7 of which have a maximum age $\leq$ 3 Myr.

The positions of these clusters in UBVI space are also consistent with older clusters. Figure~\ref{fig:CCs} displays these objects in (U-B) vs. (V-I) space as bold, red Xs in the top, middle panel, where we find all 10 of the older-appearing clusters down and to the right of the 10 Myr point in the SSP model, amongst the clusters with SED ages $>$ 10 Myr (gray points). Here we can see that the SED-fitting algorithm is determining these clusters to be young and highly reddened, as indicated by how far they must be traced in the reverse direction of the reddening vector (up and to the left) to reach the youngest end of SSP model, rather than tracing them to the nearest point on the model, which would result in older ages and smaller E(B-V). As noted in Sections~\ref{sec:section2} and \ref{sec:section72}, all of the clusters within a galaxy are fitted assuming the present-day metallicity of its young stellar population, and thus these poor estimates may be due to the fact that these older clusters are fit to an incongruent metallicity model. 

Overall, we find that most of these massive clusters with SED ages $\leq$ 3 Myr and without H$\alpha$ are consistent with globular clusters, and thus indicate that their ages and E(B-V) are poorly estimated, in agreement with \citet{WHITMORE20} and \citet{TURNER21}. While we impose an age cut for these individual inspections, \citet{WHITMORE20} find that there are likely to be several times more globular clusters with older, yet still poorly estimated SED ages, and thus the 12 objects studied here serve as a lower limit for clusters with underestimated ages. The statistical properties of all massive clusters, including these apparently mismeasured ones, are examined in Section~\ref{sec:section44}.

\subsection{\textbf{"Old" Massive Clusters with H$\alpha$}}
\label{sec:OMCs}

The second subset of clusters we examine are those that are massive ($\geq$ 5000 M$_{\odot}$), with H$\alpha$ emission (i.e. either concentrated or partially exposed H$\alpha$ morphologies), and which have SED ages $>$ 10 Myr. As listed in Table \ref{tab:YMCs+OMCs}, there are a total of 34 such cases in the sample, which account for 33$\%$ of all massive clusters with H$\alpha$. Older objects with H$\alpha$ are of interest because we generally expect clusters $>$ 10 Myr to not be producing ionizing photons, 
as massive O-stars 
will have died within this timeframe.

Upon visual inspection of these objects, we find that many appear to display characteristics consistent with young clusters. As seen in the RGB postage stamps (Figure~\ref{fig:OMCs_RGB}) 
for each cluster, there are nearby blue stars within the 75 pc radius of each of the postage stamps, signifying that they share a local space with a young stellar population. This is in addition to the clear presence of H$\alpha$ emission, which is produced by young, ionizing stars. For the clusters which lack strong blue emission (e.g. Object 207), dust could be responsible for obscuring this light, however we would need additional data, such as CO (2-1) maps from ALMA (e.g. \citealt{LEROY21}), to confirm.

\begin{figure*}
\includegraphics[width =0.85\textwidth]{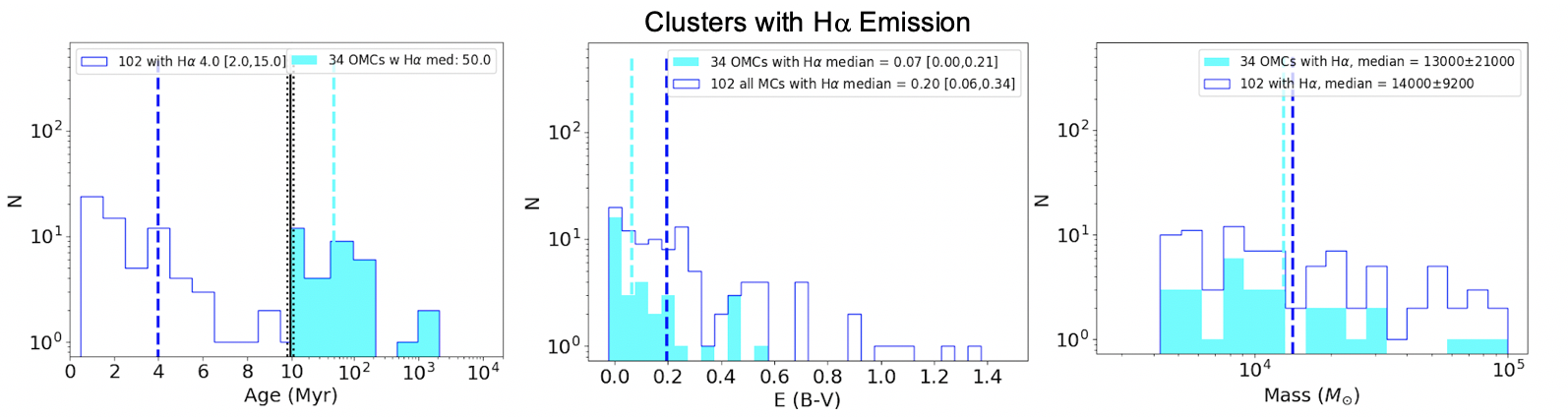}
\caption{Age, E(B-V), and mass histograms of massive ($\geq$ 5000 M$_{\odot}$) clusters with H$\alpha$ (i.e. with either concentrated or partially exposed H$\alpha$ morphologies; labeled as "OMCs"). The dark blue outline represents the distribution of all of these clusters while the cyan-filled distribution represents the oldest (SED age $>$ 10 Myr) massive clusters with H$\alpha$. The dashed, vertical dark blue and cyan lines represent the medians of all and old massive clusters with H$\alpha$, respectively. Note in the leftmost plot that the scale on the x-axis transitions from linear to logarithmic at 10 Myr.}
\label{fig:OMCs_Hist}
\end{figure*}

\begin{figure*}
\includegraphics[width =0.85\textwidth]{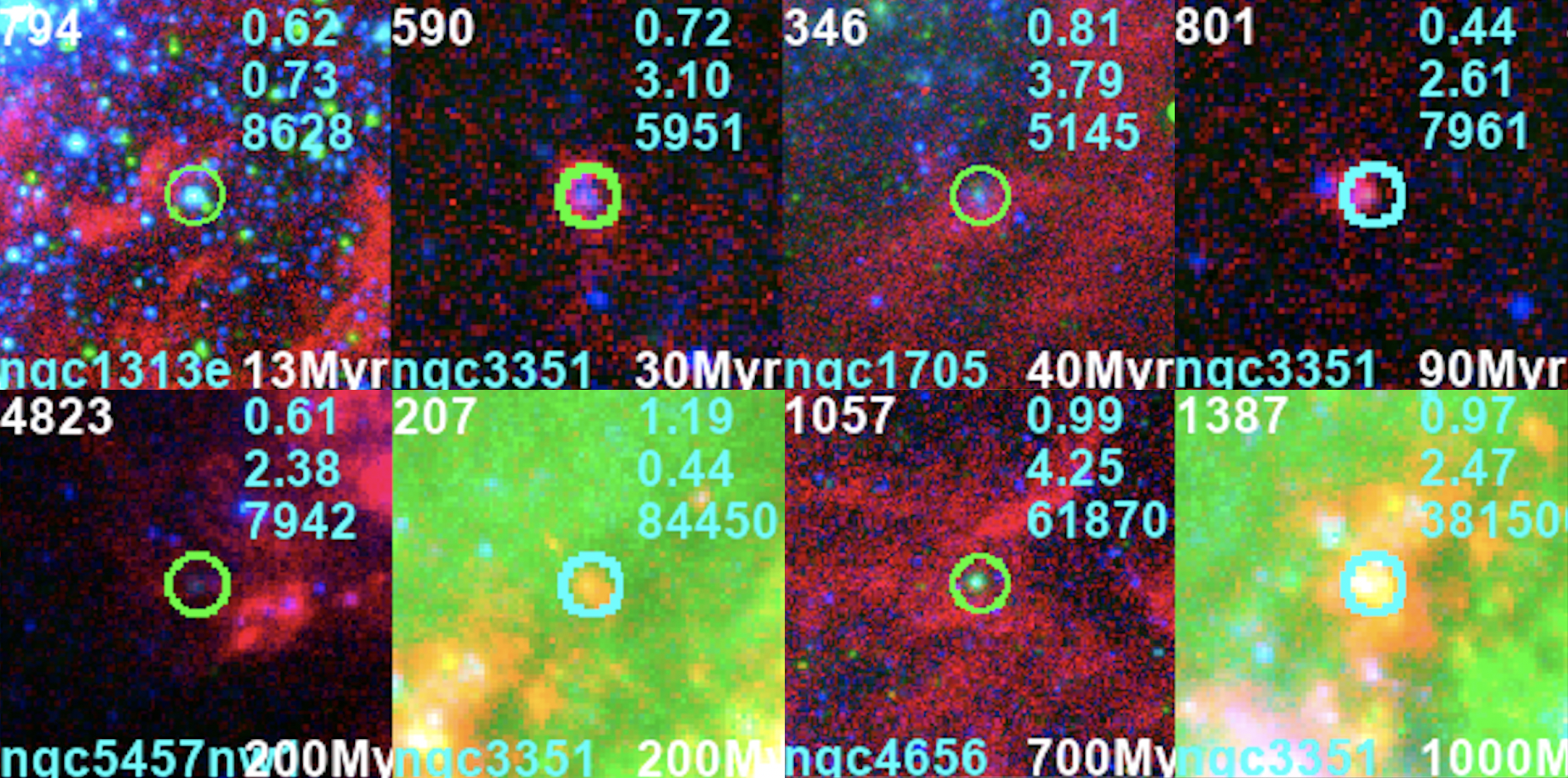}
\caption{150 pc $\times$ 150 pc RGB postage stamps for each of the 8 massive, isolated clusters with H$\alpha$ morphology classifications as either concentrated or partially exposed. Object ID is in the upper-left, field in the lower-left, and SED age in the lower-right of each stamp. In the upper-right corner of each stamp are the E(B-V), $\chi^2$ fit, and mass, from top to bottom. The bright backgrounds of Objects 207 and 1387 are due to their location at the center of NGC 3351.}
\label{fig:OMCs_RGB}
\end{figure*}

The distribution of properties for these clusters suggest that the potential age overestimation for these clusters is again due to the age-extinction degeneracy. Figure~\ref{fig:OMCs_Hist} highlights the age, E(B-V), and mass distributions of these clusters with H$\alpha$ and SED ages $>$ 10 Myr (cyan-filled) relative to the distributions for all massive clusters with H$\alpha$ regardless of SED age (blue-outlines). While there is no clear correlation with the stellar mass of these clusters, we find that the median SED age of the "older" clusters (50 Myr) is obviously well-beyond the median SED age of all massive clusters with H$\alpha$ (4 Myr). Ionizing flux can be sustained beyond 10 Myr in massive binary systems (\citealt{WOFFORD16}, \citealt{ELDRIDGE17}), however we also see that these clusters have relatively low E(B-V), with a median reddening $\sim$3x less than the median reddening of all massive clusters with H$\alpha$ (0.07 vs. 0.20 mag). Despite the clear presence of H$\alpha$, 10 of these clusters (30$\%$) have best-fit E(B-V) = 0.0 mag. Thus we find that the ages of these clusters appear to be overestimated while their E(B-V) is underestimated. Of concern is that the SED errors are limited, as was also observed amongst the "young" clusters without H$\alpha$ (Section~\ref{sec:YMCs}). Only 7 of these clusters (20$\%$) have minimum SED ages $\leq$ 10 Myr, none of which belong to the sample of isolated clusters displayed in Figure~\ref{fig:OMCs_RGB}, 
and only 3 have minimum ages $<$ 7 Myr. 

The age-extinction degeneracy can also be observed for these clusters in color space. Upon examination of their positions in (U-B) vs. (V-I) space (top, middle panel of Figure~\ref{fig:CCs}), where the bold blue cirlces and bold green triangles denote these "old" massive clusters with H$\alpha$, we find that they mostly occupy a region with clusters of SED ages $>$ 10 Myr (gray points). This, however, is not unexpected for clusters with these H$\alpha$ morphologies, as the presence of H$\alpha$ typically correlates with higher extinction (e.g. \citealt{HOLLYHEAD15}). The SED-fitting could theoretically trace them back to the SSP model below 10 Myr given a large enough E(B-V), as was the case for the "young" clusters without H$\alpha$ from Section~\ref{sec:YMCs} (bold red Xs), however in these cases, better SED fits were found for more proximal (i.e. older) points along the SSP model. 

Thus we observe the opposite effect of the age-extinction degeneracy amongst these "old" clusters with H$\alpha$ compared to those discussed in Section~\ref{sec:YMCs}, where the SED ages of these clusters appear to be overestimated while E(B-V) is correspondingly underestimated. Although we identify two specific cases of poor SED age-estimates in this section and Section~\ref{sec:YMCs}, we note that there are other types of questionable SED ages that the current study is not sensitive to, since they do not involve the presence or absence of H$\alpha$. The total number of bad ages for previous SED fitting studies such as \citet{WHITMORE20,TURNER21} have been found to be around 10-20\%, and thus the 46 of 1408 massive clusters (3\%) we investigate in the current study represent a lower limit. The impact of mismeasured properties on the overall statistical properties of the sample are discussed in the following subsection.

\subsection{Statistics of Massive Clusters}
\label{sec:section44}


Previous studies between star clusters and H$\alpha$ morphology (e.g. \citealt{WHITMORE11,HOLLYHEAD15}) have adopted a mass limit of 5000 M$_{\odot}$ in an attempt to reduce stochastic sampling effects such as those observed in Section~\ref{sec:section72}, where bright red point sources can mimic the effects of reddening due to dust. Even amongst these massive clusters, however, we find that the SED-fitting algorithm can still have difficulty in accurately estimating a cluster's age and reddening, as discussed in Sections~\ref{sec:YMCs} \& ~\ref{sec:OMCs}. With these considerations, it is important to examine how the statistical properties of massive clusters compare to the results from our overall sample.

Following \citet{WHITMORE11} \& \citet{HOLLYHEAD15}, we define our sample of massive clusters to be those that are greater than 5000 M$_{\odot}$. Upon employing this mass limit, we find $\sim$170 of these massive clusters to have SED ages $\leq$ 10, which represents an increase in sample size by a factor of $\sim$4 compared to our previous analysis \citep{HANNON19}. 

For clusters with SED ages $\leq$ 10 Myr, the median age of massive clusters are consistent with the overall sample for each of the H$\alpha$ morphological classes. Isolated clusters with concentrated H$\alpha$, partially exposed H$\alpha$, and no emission have median ages of 1.5 [1.0, 3.8] Myr, 3.0 [1.8, 5.0] Myr, and 7.0 [2.0, 9.0] Myr, respectively. Non-isolated clusters have median ages of 2.0 [1.0, 4.0] Myr, 2.0 [1.0, 4.0] Myr, and 8.0 [4.0, 9.0] Myr for clusters with concentrated H$\alpha$, partially exposed H$\alpha$, and no emission, respectively. 

If we include clusters with SED ages $>$ 10 Myr as highlighted in Section~\ref{sec:OMCs}, we find that those with concentrated H$\alpha$ overall retain a median SED age of $\sim$2.0 Myr regardless of the SED model used, which is consistent with our prior results. For clusters with partially exposed H$\alpha$, we find that the isolated sample has a median SED age ranging from 2.0-5.0 Myr, depending on the adopted SED model, which is also roughly consistent with the results of our overall sample. The non-isolated sample shows a larger range in median SED age (2.0-13.5 Myr), although only one of the six models (Padova stellar evolution; Milky Way extinction) produces a median SED age $>$ 4 Myr for this group of clusters, and thus we mostly find good agreement between the SED ages of these massive clusters and those of the overall sample.

As for the reddening of clusters, we find that E(B-V) is generally larger for the massive cluster sample. E(B-V) for massive clusters with partially exposed H$\alpha$ ($\sim$0.25 mag) is about five times larger than for the overall sample, regardless of the adopted SED model and whether or not the clusters are isolated. Massive clusters with concentrated H$\alpha$ show higher variance in median E(B-V) between SED models due to the fact that there are only 4 such isolated clusters, but if we combine the isolated \& non-isolated clusters for a larger sample (N$\approx$40), the median E(B-V) based on each of the models is $\sim$0.25 mag, or about 70\% larger than for the overall sample. We also find that massive clusters without H$\alpha$ have a median E(B-V) ($\sim$0.75) about four times larger than for the overall sample ($\sim$0.20). 

Whereas the overall sample shows that clusters with concentrated H$\alpha$ have $\sim$3 times larger E(B-V) than clusters with partially exposed H$\alpha$, we find that the E(B-V)s are comparable between the two classes for the massive cluster sample, which may further indicate the similarity between the two H$\alpha$ classes. Additionally, the relative increase in E(B-V) for both of these H$\alpha$ classes could be explained by the fact that massive clusters are more likely to remain detectable with larger attenuation. For the clusters without H$\alpha$, we find in both the massive sample and the overall sample that their E(B-V) is overestimated. While we found bright red point sources to be responsible for their overestimated E(B-V) in Section~\ref{sec:section72}, we find that globular clusters are likely responsible for the overestimated E(B-V) in the massive sample, as discussed in Section~\ref{sec:YMCs}, and in agreement with similar star cluster studies \citep{WHITMORE20,TURNER21}. 

Further, the relative fraction of massive clusters in each H$\alpha$ class is also consistent with our overall sample. For clusters with SED ages $\leq$ 10 Myr, the percent of clusters with concentrated and partially exposed H$\alpha$ classes are $\sim$25\% and $\sim$15-20\%, respectively, which imply that the onset of gas clearing begins early ($\lesssim$ 3 Myr) and takes place over a short interval (1-2 Myr). Considering the clusters with apparently mismeasured SED ages, if we exclude the clusters without H$\alpha$ which appear much older than their SED ages (Sections~\ref{sec:YMCs}) and include those with H$\alpha$ and overestimated SED ages (\ref{sec:OMCs}), the resultant percentages of isolated clusters with concentrated and partially exposed H$\alpha$ morphologies is roughly consistent with the above values -- $\sim$15\% \& $\sim$23\%, respectively. 

Thus we find in our analysis of the massive star cluster sample that despite the presence of objects with apparently mismeasured SED ages (Sections~\ref{sec:YMCs} \& \ref{sec:OMCs}), we still find overall good agreement between the statistical ages of the massive cluster sample and those of the total sample, and therefore our reported gas clearing timescales are upheld.


\section{Uncertainties}
\label{sec:section5}

\subsection{LEGUS vs. CIGALE SED-fitting}
\label{sec:cigale}

In our statistical presentation of star cluster properties, we have found that there are sometimes small or even no differences between the median ages, E(B-V)s, and masses of clusters within each H$\alpha$ morphological class. Especially considering the the difficulty that SED-fitting may have with the age-extinction degeneracy (Section~\ref{sec:section73}), our confidence level in these star cluster properties and their distinctions is pertinent to drawing meaningful conclusions from their comparisons.

To this point, we have only considered the best-fit properties (those that minimize $\chi^2$) derived from a single SED-fitting algorithm as outlined in Section~\ref{sec:section2}. As one way of testing the robustness of these properties, we follow the method of \citet{TURNER21} in which we use an independent SED-fitting algorithm, Code Investigating GALaxy Emission (CIGALE; \citealt{BURGARELLA05,NOLL09,BOQUIEN19}), to produce new catalogs of properties for our sample of star clusters. For valid comparison, CIGALE allows us to input the same assumptions made by LEGUS, namely Yggdrasil SSP models, a Kroupa IMF, Milky Way extinction, and a covering fraction of 0.5.

With these same input parameters, we produce two new star cluster catalogs with CIGALE. For the first of these, we simply use the LEGUS broadband photometry as direct input for CIGALE to perform the SED-fitting. From this fitting, CIGALE will output not only the desired star cluster properties, but also the photometry of the best model fit for each cluster. For the second catalog, fluxes for each cluster are randomly chosen from Gaussian distributions centered on its best model fit, where the standard deviations are based on the individual photometric errors. Once all fluxes are thus determined, they are then fed back into CIGALE to produce new star cluster properties. In this manner, we have 1) a direct comparison of SED-fitting algorithms using the same photometry, and 2) a set of “mock clusters” for comparison. To note, the age grids of sub-solar metallicity models in CIGALE are truncated relative to the LEGUS models, thus our re-fitting is performed for the 13 fields for which LEGUS assumes solar metallicity (see Table~\ref{tab:table1}). This accounts for 2787 of the 3757 (73$\%$) star clusters of our sample, for which we present the following analysis.

\begin{figure*}
\includegraphics[width =1.0\textwidth]{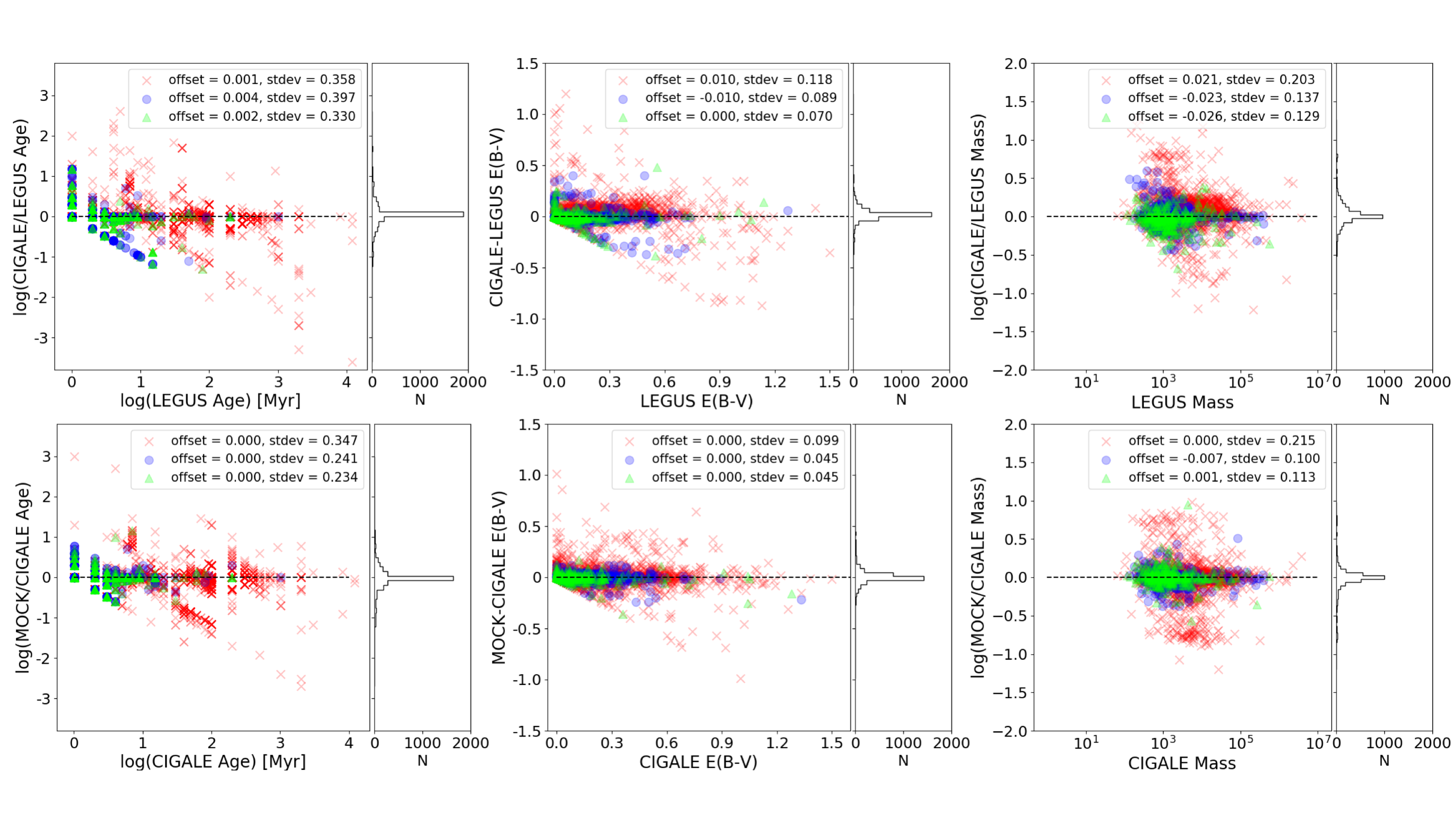}
\caption{Comparison of star cluster properties (age, E(B-V), and mass in columns from left to right) between three SED-fitting methods. The top row compares LEGUS results (as reported throughout this paper) with the results of CIGALE's SED-fitting of the same LEGUS photometry, while the bottom row compares these initial CIGALE results with a set of Gaussian-determined "mock clusters" also fitted with CIGALE. While there appears to be considerable scatter amongst these points, the histograms provided to the right of each plot indicate that the vast majority of clusters have consistent properties between the three methods. The median offset and standard deviation for each property, separated by H$\alpha$ morphology, are shown in the legend of each plot and highlight the small overall differences between methods.}
\label{fig:CIGALE_Comp}
\end{figure*}

Overall, the median difference in each of the cluster properties between the catalogs is small, although we note that there is considerable scatter.  Figure~\ref{fig:CIGALE_Comp} displays the plots comparing each of the three star cluster properties according to the three catalogs. The top row shows the comparisons between LEGUS fitting and the initial CIGALE fitting while the bottom row compares the initial CIGALE fitting to the mock cluster fitting. To the right of each of these plots is a histogram corresponding to the true distribution of points, which serves to highlight that although some clusters show large (e.g. up to $\sim$2 dex in age) disagreements between the star cluster catalogs, the vast majority of clusters show small or no discrepancies. As a result, we find that the overall differences between the catalogs are quite small: the median ratio of cluster ages is less than 0.005 dex for each H$\alpha$ morphological class and for each catalog comparison, while the median ratio of cluster mass is less than 0.05 dex for each of the comparisons. The median difference in E(B-V) is no more than 0.02 mag across each H$\alpha$ class and catalog, and the majority of these comparisons have no difference in median E(B-V). These results are consistent with \citet{TURNER21}, who found median age and mass ratios of 0.001 and 0.003 dex, respectively, when comparing LEGUS \& CIGALE catalogs for the 292 clusters of NGC 3351, one of the galaxies in our sample. While these median differences are small, scatter is considerable: the standard deviation in the sample of properties are $\sim$0.35, $\sim$0.10, and $\sim$0.20 for age, E(B-V), and mass, respectively.

The observed scatter can in part be attributed to systematic effects. The diagonal striping that can be seen in the age comparisons in the left column of Figure~\ref{fig:CIGALE_Comp} is due to the discrete nature of the allowed ages in the LEGUS and CIGALE models (i.e., 1, 2, 3, 4 … Myr), as can be seen by the discrete vertical lines. For example, in the upper left panel, the lowest diagonal represents values of LEGUS = 1, 2, 3, 4, … Myr and CIGALE = 1 Myr. The next higher diagonal stripe represents LEGUS = 1, 2, 3, 4 … Myr and CIGALE = 2 Myr. The discrete ages also produce systematic shifts in the ages and E(B-V)s of clusters at the extreme ends of our fixed grids, where a cluster with LEGUS age = 1 Myr must have a CIGALE age $\geq$ 1 Myr, which contribute to the standard deviations shown in the legends of Figure~\ref{fig:CIGALE_Comp}. 

These systematic effects as well as random uncertainties are reflected in the standard deviations, where we glean that cluster ages are consistent between the SED-fitting algorithms to within $\sim$0.35 dex or about a factor of 2, as was also found in previous studies such as \citet{WHITMORE20}. While we do find a few more extreme discrepancies as high as 3 dex, these objects are mostly clusters without H$\alpha$, which include globular clusters (see also \citet{WHITMORE20}. These objects are the subject of discussion in Section~\ref{sec:YMCs}.

Despite the notable scatter in our comparisons, we still see very small overall shifts in the statistical properties of star clusters, and do not affect the results we presented in Section~\ref{sec:section4}. The median SED ages of clusters with concentrated (1-2 Myr), partially exposed (2-3 Myr), and absent H$\alpha$ (4-5 Myr), as given by the two CIGALE-based catalogs, are consistent with what we have found based on the LEGUS catalogs. This consistency similarly holds for cluster reddening, where the median E(B-V)s of clusters with concentrated, partially exposed, and absent H$\alpha$ are 0.15, 0.04-0.05, and 0.14-0.15 mag, respectively. 

Overall, this exercise demonstrates that the SED-fitting algorithms used by PHANGS-HST and LEGUS are consistent, however this does not mean that the age estimates themselves are all correct, only that they are reproducible. In fact, the clusters with apparently mismeasured ages which we investigate in Sections~\ref{sec:YMCs} \& \ref{sec:OMCs} have similarly poor age estimates when fitted with CIGALE instead. As discussed in Section \ref{sec:section72} above, and in more detail in \citet{WHITMORE20} \& \citet{TURNER21}, the SED ages appear to be underestimated in as much as $\sim$25\% of the older cluster population in favor of overestimated E(B-V), based on the number of clusters without H$\alpha$ and young age estimates either due to red point sources or an overall red color. For a more detailed analysis of the SED-fitting algorithms presented in this Section for the star clusters of one of our galaxies (NGC 3351), we refer the reader to \citet{TURNER21}.

\subsection{Resolution}
\label{sec:resolution}

The galaxies in our sample span a range of distances from $\sim$3--10 Mpc. Because the nearest galaxies in the sample will have $\sim$3x better physical resolution than the most distant ones, an important question to address is whether the H$\alpha$ morphological classifications are dependent on the resolution of our images. To test this, we degrade H$\alpha$ imaging of our nearest galaxy (NGC 7793; 3.44 Mpc) such that it appears as though we observed it at the distance of our furthest galaxy (NGC 3351; 10.0 Mpc), and then reclassify each of the clusters according to the morphology of their degraded H$\alpha$ emission. To further examine the limits of this type of morphological analysis, we also perform reclassifications for H$\alpha$ reprojected to a distance of 20.0 Mpc as well as for ground-based H$\alpha$ imaging.

The first step in the degradation process is to smooth the image to the appropriate full width at half maximum (FWHM), which we do by convolving the image with a 2D Gaussian function with a standard deviation corresponding to the difference in quadrature between the FWHM of the original image and the expected FWHM of the new image. In order to retain a consistent pixel scale, we then reproject the image onto a smaller grid to match that of the galaxy at a projected distance of 10.0 or 20.0 Mpc. At this point it is important to subtract the background from the image, as the previous steps produce a smoothing effect that reduces noise. With the background subtracted, we scale down the signal of the image by a factor of the relative distance squared to obey the inverse square law. Next, we reintroduce the original background to the image. We reproduce this background by drawing from a Gaussian distribution such that the final image’s background level and its standard deviation match those of the original image. Finally, we introduce slight pixel-to-pixel correlation by convolving the image with a compact Gaussian. 

\begin{figure*}
\includegraphics[width =1.0\textwidth]{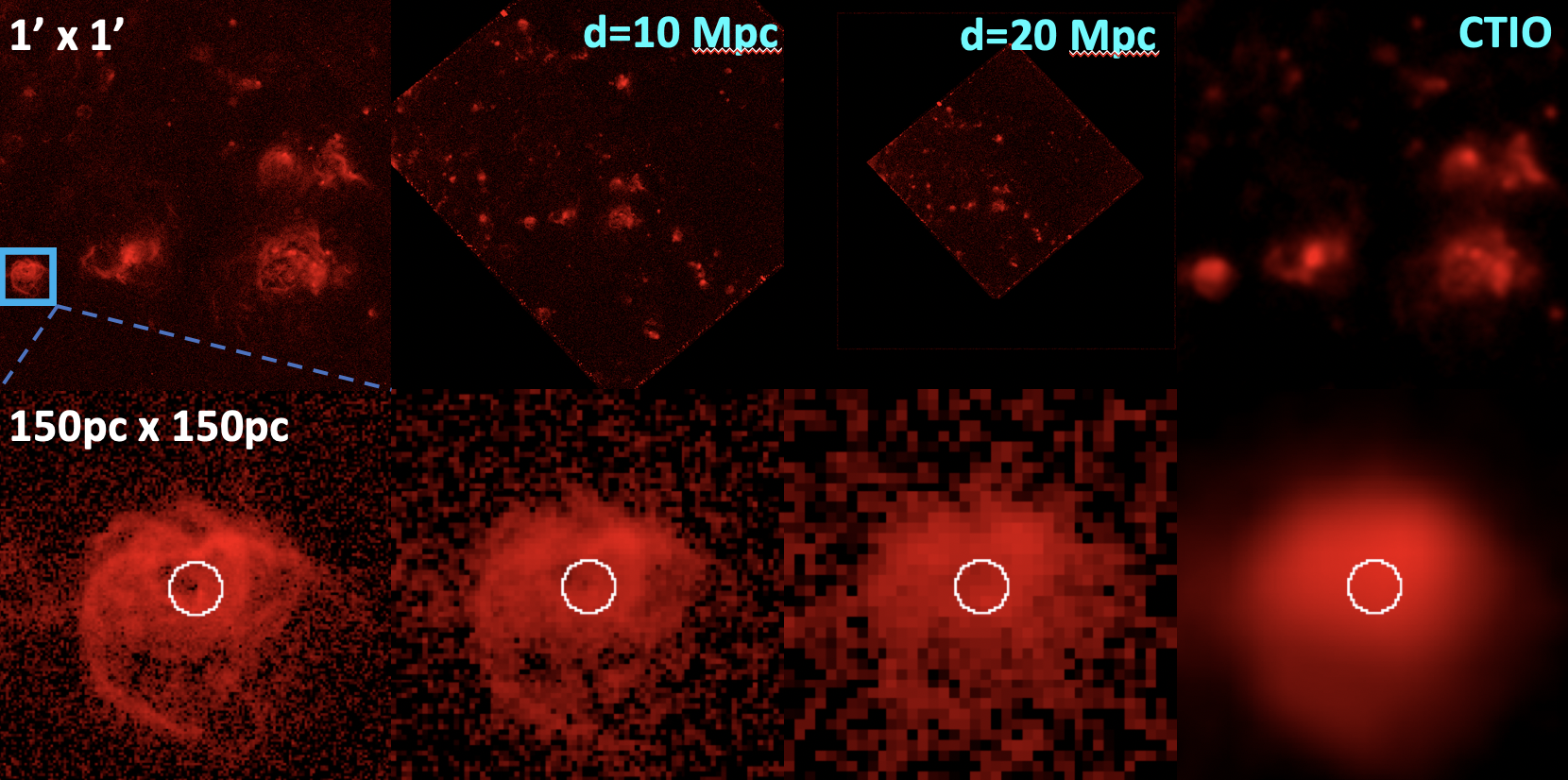}
\caption{Continuum-subtracted H$\alpha$ images of NGC 7793W with varying physical resolution. From left to right, the columns display 1) images of the original \textit{HST} data from LEGUS ($\sim$1.3 pc resolution), 2) the \textit{HST} data degraded to resemble the data at a distance of 10.0 Mpc ($\sim$4.6 pc resolution), 3) the \textit{HST} data degraded to resemble the data at a distance of 20.0 Mpc ($\sim$9.2 pc resolution), and 4) ground-based H$\alpha$ taken with the CTIO 1.5m telescope as part of the SINGS program (\citealt{KENNICUTT03}; $\sim$16.7 pc resolution). The top row shows images of the galaxy at the same angular scale (1$\arcmin$ $\times$ 1$\arcmin$) and the bottom row shows an example an example HII region at the same physical scale (150 $\times$ 150 pc).}
\label{fig:Resolution_Comps}
\end{figure*}

Figure~\ref{fig:Resolution_Comps} illustrates the results of such image degradation in comparison to the original \textit{HST} image for NGC 7793W as well as overlapping ground-based H$\alpha$ taken with the Cerro Tololo Inter-American Observatory (CTIO) 1.5m telescope as part of the Spitzer Infrared Nearby Galaxies Survey (SINGS) program \citep{KENNICUTT03}. The top row of images share an angular scale (1$\arcmin$ $\times$  1$\arcmin$) while the bottom row shows an example HII region at a shared physical scale of 150pc $\times$ 150pc. Compared to the original H$\alpha$ image (leftmost column), we see in the top row that the two degraded images (middle columns) appear to have a similar background level, while the brightest regions become expectedly smaller and fainter. When we zoom into the HII region shown in the bottom row, we note that the clearing at the center of the region, which is obvious in the original image, becomes smoothed over as the image degrades. The central dip in flux remains visible at a projected distance of 10 Mpc, but appears almost imperceptible at a projected distance of 20 Mpc, and is completely undetectable in the CTIO image (right column). On its short-axis, the cleared region at the center of bubble is $\sim$10 pixels in diameter in the original image, which is commonly observed for these bubble-like morphologies in this analysis. At a projected distance of 20 Mpc, these same areas of clearing would be less than 2 pixels wide, hence why it appears that this distance represents a natural limit to identifying these structures. The CTIO image has an approximate angular resolution of 1$\arcsec$, which is $>$ 10 worse than the resolution of the \textit{HST} image ($\sim$0.08$\arcsec$) and is thus simply unable to resolve structures of this scale.
  
\begin{figure*}
\includegraphics[width =1.0\textwidth]{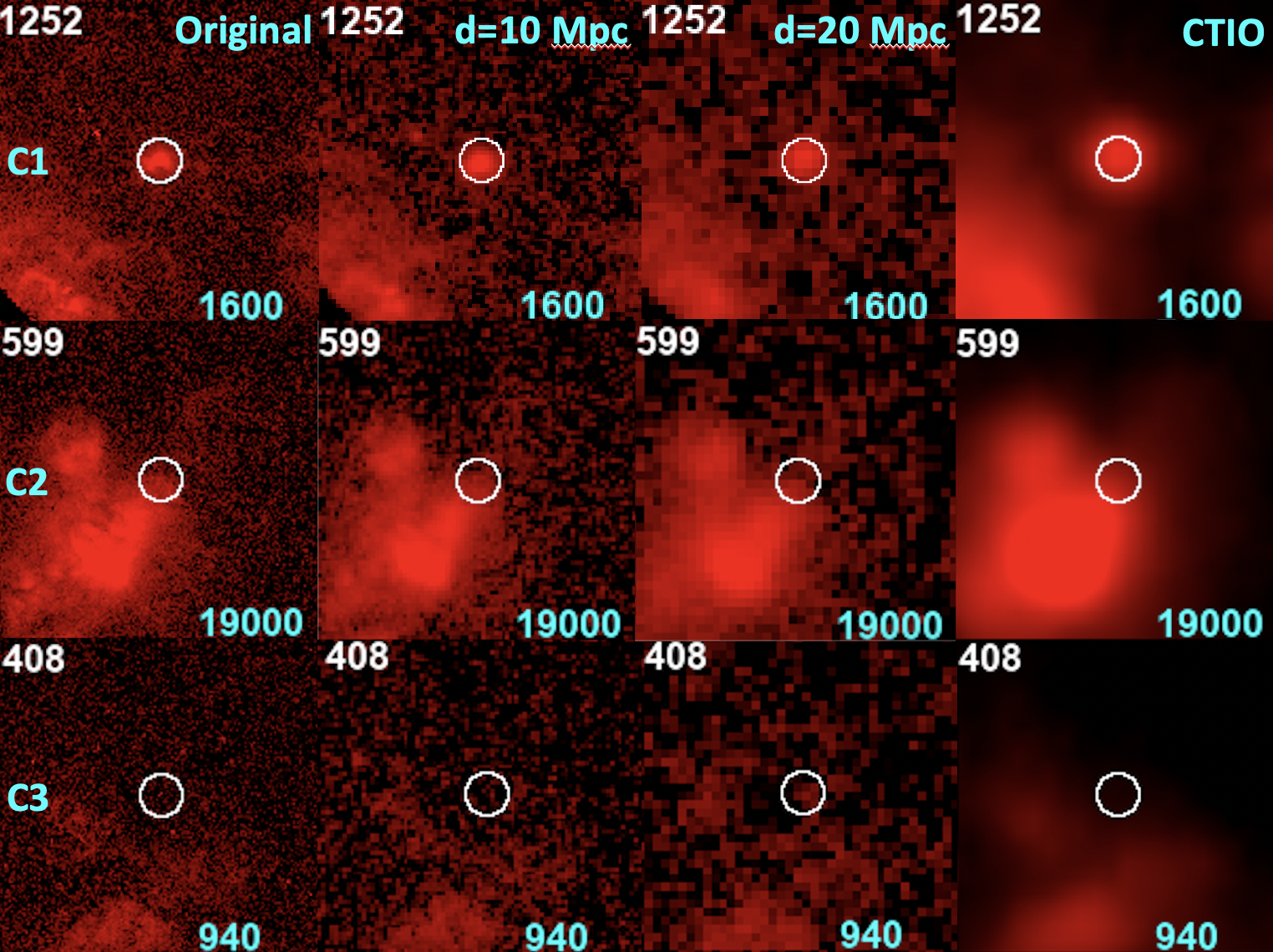}
\caption{150 pc $\times$ 150 pc postage stamps of continuum-subtracted H$\alpha$ comparing the original \textit{HST} images of three NGC 7793 clusters (first column) with their images projected to 10.0 Mpc (second column), projected to 20.0 Mpc (third column), and with the CTIO image (fourth column). The top, middle, and bottom rows show an example class 1, 2, and 3 H$\alpha$ morphology, respectively, which received the same classification according to all four images. Stamps include the Object ID in the upper-left and stellar mass in the lower-right.}
\label{fig:Resolution_Comps_123}
\end{figure*}

Once all of the images are produced, we reclassify all 221 clusters within the NGC 7793W field according to the H$\alpha$ emission from each image. The classification procedure is similar to that outlined in Section~\ref{sec:section2}: we create 150 pc $\times$ 150 pc postage stamps centered on each cluster from the LEGUS catalog and assemble the stamps into a single collage ordered by their best-fit SED age, and then into separate collages based on their initial H$\alpha$ classification. However, rather than using RGB images to aid in classification, here we simply use the continuum-subtracted H$\alpha$ to determine each cluster's H$\alpha$ morphology. Examples of H$\alpha$ morphologies which receive the same classification according to all of the images are shown in Figure~\ref{fig:Resolution_Comps_123}.

Table~\ref{tab:DegradedClasses} displays the results of the reclassifications in terms of what percent of the original H$\alpha$ classes are reclassified as a class 1, 2, or 3 (1 = concentrated, 2 = partially exposed, 3 = no emission) according to the each of the new images. Overall, we find relatively good agreement ($>$ 74\%) between the original classes and those determined using the d = 10.0 Mpc image, and we note that this agreement decreases with decreasing resolution, as expected. We also find that the H$\alpha$ morphologies which receive a new class are almost exclusively given a more concentrated classification (e.g. a partially exposed morphology gets reclassified to a concentrated morphology), due to the smoothing effect discussed earlier, regarding the example HII region in Figure~\ref{fig:Resolution_Comps}.  

\begin{table*}
\begin{tabular}{lcccccccccc}
\multicolumn{11}{c}{\textbf{Agreement Fractions for H$\alpha$ Reclassifications}} \\
\hline
        &     & \multicolumn{3}{c}{d = 10.0 Mpc ($\sim$4.6 pc)}                       & \multicolumn{3}{c}{d = 20.0 Mpc ($\sim$9.2 pc)}                       & \multicolumn{3}{c}{SINGS (CTIO; $\sim$16.7 pc)}                     \\
\hline
\hline
        & N   & Class 1          & Class 2         & Class 3         & Class 1          & Class 2         & Class 3         & Class 1          & Class 2         & Class 3         \\
\hline
LEGUS Class 1 & 28  & \textbf{100.0\%} & 0.0\%           & 0.0\%           & \textbf{100.0\%} & 0.0\%           & 0.0\%           & \textbf{100.0\%} & 0.0\%           & 0.0\%           \\
\hline
LEGUS Class 2 & 27  & 22.2\%           & \textbf{74.1\%} & 3.7\%           & 29.6\%           & \textbf{63.0\%} & 7.4\%           & 74.1\%           & \textbf{25.9\%} & 0.0\%           \\
\hline
LEGUS Class 3 & 156 & 0.0\%            & 4.5\%           & \textbf{95.5\%} & 0.0\%            & 5.8\%           & \textbf{94.2\%} & 0.0\%            & 14.1\%          & \textbf{85.6\%} \\
\hline
\hline
\end{tabular}
\caption{Comparison between H$\alpha$ morphological classifications for NGC 7793W based on the original \textit{HST} H$\alpha$ image (LEGUS Classes) and 1) \textit{HST} H$\alpha$ reprojected to a distance of 10.0 Mpc, 2) \textit{HST} H$\alpha$ reprojected to d = 20.0 Mpc, and 3) ground-based CTIO data from the SINGS program \citep{KENNICUTT03}. The physical resolution for each is also included, compared to $\sim$1.3 pc with the original image. Rows display the number of each of the original classifications and include the percentage of those that were reclassified as a class 1, 2, or 3 (1 = concentrated, 2 = partially exposed, 3 = no emission) according to each of the three new images. As physical resolution decreases, the recovery rate of the original classes decreases, as expected. This is due to the fact that the cleared regions immediately surrounding a cluster are often smoothed over, which results in a more concentrated classification, e.g. an HII region originally identified as class 2 appears instead as a class 1 according to the lower resolution image.}
\label{tab:DegradedClasses}
\end{table*}

The H$\alpha$ class most affected by the reclassification process is the partially exposed class of objects, due to the fact that it requires the resolution of the smallest structures such as bubbles and filaments. When the galaxy is projected out to a distance of 10.0 Mpc, 74\% of the partially exposed H$\alpha$ morphologies are retained, while 63\% are retained when projected out to 20.0 Mpc, and only 26\% are retained when using the ground-based imaging. The remaining clusters are almost all reclassified as having concentrated H$\alpha$, as the lower resolution smooths over small clearings, however a couple of HII regions become too faint for observation at their further projected distances and as such are reclassified as having no emission. Additionally, all of the clusters classified as having partially exposed H$\alpha$ according to the CTIO image are found at the edges of crowded regions (having neighboring clusters within 75 pc; see Object 599 in Figure~\ref{fig:Resolution_Comps_123}); no clear bubble structures, as seen in the original image of Object 2099 (Figure~\ref{fig:Resolution_Comps}), are identified with the ground-based H$\alpha$ image.

Better agreement is found amongst clusters without H$\alpha$, as $\lesssim$ 10\% of these objects are given a new classification based on the lower-resolution images. Each of these objects are reclassified as partially exposed, as more distant H$\alpha$ is smoothed such that it appears closer to the cluster than originally observed. This effect can be seen for Object 408 in Figure~\ref{fig:Resolution_Comps_123}, however in this case, the smoothed H$\alpha$ remains distinct enough to retain the cluster’s H$\alpha$ class 3 label. Regardless of the H$\alpha$ image used, all of the clusters with an originally-determined concentrated morphology retained their classification, as smoothing does not affect this classification and the regions remained sufficiently bright for observation (e.g. Obejct 1252 in Figure~\ref{fig:Resolution_Comps_123}). 

Upon these reclassifications, it is important to examine their effects on our statistical results of cluster properties. In Table~\ref{tab:DegradedClassStats}, we show the median age (top) and reddening (bottom) for clusters with SED ages $\leq$ 10 Myr according to their respective classifications. For all of the images used to classify the H$\alpha$ morphologies, we find equivalent median ages for the three classes: 3 Myr for clusters with concentrated H$\alpha$, 3 Myr for clusters with partially exposed H$\alpha$, and 5 Myr for clusters without H$\alpha$, and the median ages of clusters with concentrated and partially exposed H$\alpha$ do not change when clusters with SED ages $>$ 10 Myr are included. The age distributions for these clusters show that the majority ($\gtrsim$ 60\%) of those which have changed classifications are between 2-5 Myr. Additionally, while we find a modest peak of clusters with concentrated H$\alpha$ at 3 Myr based on the \textit{HST} H$\alpha$, the peak becomes much more pronounced as the physical resolution decreases, which also diminishes the peak of these clusters at 1 Myr which was much more clearly observed for the entire sample in Figure~\ref{fig:AgeHist}. Without a strong peak at 1 Myr, the median age of clusters with concentrated H$\alpha$ does not change based on whether we choose an age limit of 10 Myr or 5 Myr, regardless of the image used, contrary to what is observed for the whole sample (Section~\ref{sec:section41}), though this could be an effect of the smaller sample size. Thus find even greater consistency amongst the ages of clusters with concentrated and partially exposed H$\alpha$ when the sample is reduced to a single field in NGC 7793W (median = 3 Myr). 

\begin{table*}
\begin{tabular}{lcccccccc}
\multicolumn{9}{c}{\textbf{Statistical Star Cluster Properties for H$\alpha$ Reclassifications}} \\
\hline
                  & \multicolumn{2}{c}{Original H$\alpha$} & \multicolumn{2}{c}{d = 10.0 Mpc} & \multicolumn{2}{c}{d = 20.0 Mpc} & \multicolumn{2}{c}{SINGS (CTIO)} \\
\hline
\hline
                  & N         & Median Age (Myr)           & N      & Median Age (Myr)        & N      & Median Age (Myr)        & N      & Median Age (Myr)        \\
\hline
Concentrated      & 28        & 3.0 {[}1.8,3.2{]}          & 34     & 3.0 {[}2.0,3.0{]}       & 39     & 3.0 {[}2.0,3.0{]}       & 48     & 3.0 {[}2.0,3.0{]}       \\
\hline
Partially Exposed & 25        & 3.0 {[}2.0,3.0{]}          & 23     & 3.0 {[}2.0,3.5{]}       & 19     & 3.0 {[}3.0,3.5{]}       & 22     & 3.0 {[}2.2,4.8{]}       \\
\hline
No Emission       & 76        & 5.0 {[}3.0,7.0{]}          & 72     & 5.0 {[}3.0,7.0{]}       & 71     & 5.0 {[}3.0,7.0{]}       & 59     & 5.0 {[}3.0,7.0{]}       \\
\hline
\hline
                  & N         & Median E(B-V)              & N      & Median E(B-V)           & N      & Median E(B-V)           & N      & Median E(B-V)           \\
\hline
Concentrated      & 28        & 0.12 {[}0.09,0.26{]}       & 34     & 0.11 {[}0.06,0.23{]}    & 39     & 0.11 {[}0.05,0.20{]}    & 48     & 0.10 {[}0.05,0.17{]}    \\
\hline
Partially Exposed & 25        & 0.06 {[}0.00,0.13{]}       & 23     & 0.07 {[}0.01,0.13{]}    & 19     & 0.10 {[}0.05,0.14{]}    & 22     & 0.11 {[}0.07,0.14{]}    \\
\hline
No Emission       & 76        & 0.10 {[}0.06,0.18{]}       & 72     & 0.11 {[}0.06,0.18{]}    & 71     & 0.11 {[}0.06,0.18{]}    & 59     & 0.11 {[}0.06,0.18{]}  \\
\hline
\hline
\end{tabular}
\caption{Comparison of the age (top) and E(B-V) (bottom) statistics of clusters whose H$\alpha$ morphologies have been determined using, from left to right: 1) the original \textit{HST} H$\alpha$ image, 2) \textit{HST} H$\alpha$ reprojected to a distance of 10.0 Mpc, 3) \textit{HST} H$\alpha$ reprojected to d = 20.0 Mpc, and 4) ground-based CTIO data from the SINGS program \citep{KENNICUTT03}. Statistics are shown for clusters with SED ages $\leq$ 10 Myr, however these results are largely unaffected when the age cut is removed (except for clusters without H$\alpha$, whose median age expectedly increases with the inclusion of older clusters).}
\label{tab:DegradedClassStats}
\end{table*}

Additionally, if we examine the relative fraction of H$\alpha$ morphologies with cluster SED ages $\leq$ 10 Myr as done in Section~\ref{sec:section41}, we find that the percent of clusters with concentrated H$\alpha$ notably increases as resolution decreases. The classifications based on the original H$\alpha$ and the two degraded images show that 20-30\% of clusters have concentrated H$\alpha$ morphologies, which implies that gas clearing begins at 2-3 Myr, consistent with our previous findings. The classifications based on the SINGS data, however, show that almost 40\% of these clusters have concentrated H$\alpha$, which would indicate that clearing begins at an age ($\sim$4 Myr) older than previously discussed, and would also make a crucial change to our conclusions regarding the feedback mechanisms responsible for clearing, as this older age would then allow for supernovae to have already begun igniting and thus take emphasis away from other feedback mechanisms such as stellar winds and radiation pressure.

For cluster reddening, we find very small differences ($\sim$0.01 mag) between the original classifications and the 10.0 Mpc reprojection, though we also observe a notable converging of E(B-V) amongst the three classes as the physical resolution decreases. This convergence occurs due to clusters originally classified with partially exposed H$\alpha$ losing clusters with low E(B-V) to the concentrated class of objects while gaining clusters with relatively higher E(B-V) which were originally classified as having no emission.

Overall, we find good agreement in both the visual H$\alpha$ classifications and statistical properties of clusters in NGC 7793W at the original physical resolution and when the galaxy is projected out to a distance of 10.0 Mpc. This indicates that our analysis of H$\alpha$ morphologies across a distance range of $\sim$3--10 Mpc is robust to resolution effects. 
We also project the galaxy out to a distance of 20.0 Mpc, where we appear to be around the limit for the resolution of fine bubble and filamentary structures commonly used in the classification of partially exposed morphologies. When we examine ground-based H$\alpha$ data, we find particularly poor agreement in identifying partially exposed H$\alpha$ morphologies, as it is impossible to resolve their common small structures, which serves to highlight the importance of \textit{HST}’s resolution capability in identifying early stages of gas clearing and thereby understanding the mechanisms responsible for it.

\section{Summary \& Conclusions}
\label{sec:section8}

To expand on our work in \citet{HANNON19}, we analyze 3757 star clusters in 21 \textit{HST} WFC3/UVIS fields covering 16 galaxies (spanning a distance range of $\sim$3--10 Mpc) from the LEGUS sample (GO-13364; PI D. Calzetti). This study includes all LEGUS galaxies with both \textit{HST} H$\alpha$ narrowband imaging (GO-13773; PI R. Chandar) and cluster catalogs containing SED-fit cluster properties \citep{ADAMO17} published as of 2022 January, plus an unpublished catalog of clusters in NGC 5457. We study all visually-identified cluster class 1, 2, and 3 objects corresponding to symmetric, asymmetric, and multi-peaked clusters, respectively. Relative to \citet{HANNON19}, the galaxy sample is increased by a factor of $\sim$4 and the star cluster sample increased by a factor of $\sim$6.
One key addition in this study is that H$\alpha$ morphology classifications have been made for all star clusters regardless of SED age while \citet{HANNON19} only considered clusters with SED ages $\leq$ 10 Myr.

By comparing different H$\alpha$ morphologies with the properties of their host star clusters, which effectively represent single-aged stellar populations, we seek to gain insight into the timescales, and thus the physical processes at work, in the clearing of a cluster’s natal gas. A multi-stage process of visual inspection is employed to classify clusters according to their H$\alpha$ morphology (concentrated, partially exposed, or no emission), and the clusters are further categorized by whether they have neighboring clusters ($\leq$ 75 pc away), which could affect the clearing timescales.

Here we present a summary of our results:

\begin{enumerate}

	\item Of the clusters with SED ages $\leq$ 10 Myr, those with concentrated H$\alpha$ have a median age of 1-2 Myr, those with partially exposed H$\alpha$ have a median age of 2-3 Myr, and those without H$\alpha$ have a median age of 3-6 Myr (ranges reflect differences resulting from various dust and stellar population models used in the SED fitting). 
	These medians represent a shift to slightly younger ($\leq$ 1 Myr) ages than was found in \citet{HANNON19}. Together with the inferred ages from the relative fraction of each H$\alpha$ morphology, they support the two main conclusions of \citet{HANNON19} and others: Firstly, the prevalence of clusters which show evidence of gas clearing (partially exposed H$\alpha$ class) and yet have SED ages younger than the possible onset of supernovae ($\leq$ 3 Myr) highlight the importance of pre-supernova gas clearing mechanisms such as stellar winds, radiation pressure, and photoionization during the first few Myr of a cluster's life (see also \citealt{CHEVANCE20}). Secondly, the age difference between clusters with concentrated and partially exposed H$\alpha$ suggests short (1-2 Myr) clearing timescales. 
    \item Clusters with concentrated H$\alpha$ and those without H$\alpha$ emission share similar median E(B-V) values ($\sim$0.2 mag). With a median cluster mass of $\sim$1000 M$_{\odot}$, we find that the reddening of clusters without H$\alpha$ are likely overestimated due to the presence of bright red stars, a result of stochastic sampling of the IMF. We also see clusters with concentrated H$\alpha$ which are found blueward of the young end of the SSP models, potentially due to additional contributions from nebular emission \citep{ZACKRISSON01}, or the presence of two or more Wolf-Rayet stars (type WC or WNE; see \citealt{OROZCO21}). These latter stochastic sampling effects instead have a bias toward bluer colors, and thus lower E(B-V).
    
    \item We find that at least 46 of 1408 (3\%) massive ($\geq$ 5000 M$_{\odot}$) clusters appear to have questionable SED ages, based on our H$\alpha$ study. These fall into two categories. The first group of anomalies consists of 12 isolated clusters with no H$\alpha$ emission, yet which have very young ($\leq$ 3 Myr) apparent SED ages. Upon visual inspection of each object’s image, 10 of the 12 appear to have features consistent with much older clusters (i.e., globular clusters), including their color, symmetry, lack of nearby young stars, lack of H$\alpha$ emission, and presence in the bulge. The second group of anomalies consists of 34 massive clusters with H$\alpha$ emission (2\% of all massive clusters), but apparent SED age estimates that are old ($>$ 10 Myr). We note that there are other types of questionable SED age estimates that the current study is not sensitive to, since they do not involve the presence or absence of H$\alpha$. Hence our estimates represent lower limits to the total number of poor age estimates (e.g., see \citet{WHITMORE20} \& \citet{TURNER21}, who find $\sim$10-20\% of clusters have poor estimates in NGC 4449 and NGC 3351, respectively).

    \item The median ratios of cluster ages and masses between our standard LEGUS catalog and results derived using CIGALE with the parameters described in \citep{TURNER21} are less than 0.005 dex and 0.05 dex, respectively, for all H$\alpha$ morphologies. For cluster E(B-V), we find the median difference to be less than 0.02 mag for all such comparisons and morphologies. These small offsets along with the fact that our statistical results (see point 1) remain unaffected indicate that our star cluster properties are robust to the adopted SED-fitting algorithm and thus support our reported timescales. 
    \item We find that the classification of H$\alpha$ morphologies does not change significantly over the distances spanned by the galaxies in this sample (3--10 Mpc). 
    For our nearest galaxy (NGC 7793W; 3.44 Mpc), the majority ($>$ 64\%) of \textit{HST}-based partially exposed classifications are recovered when we project the galaxy to be at a distance of our most distant galaxy (NGC 3351; 10.0 Mpc), and even to a distance of 20.0 Mpc, at which we appear to reach the resolution limit for our classifications. For both of these reprojections, the statistical properties of clusters according to their H$\alpha$ morphologies remain consistent with the original classifications. The use of ground-based H$\alpha$ data taken at CTIO as part of the SINGS program, however, is unfit for this analysis as it is simply unable to resolve small clearing structures necessary to identify the early stages of gas clearing, and therefore we find that the resolution capability of \textit{HST} is necessary to produce the analysis and results presented in this paper.

\subsection{Future Work}
Our analysis highlights issues due to the age-extinction degeneracy.
While we identify instances in which ages of clusters with H$\alpha$ are overestimated and vice-versa, they do not occur frequently enough to affect the median ages of the different H$\alpha$ morphological classes.

To potentially break the observed age-extinction degeneracy in the SED fitting, there are a few avenues available to pursue: 1) there exists MUSE (\textit{Multi Unit Spectroscopic Explorer}) data for fields overlapping with our sample with which we could create E(B-V) maps, as done by \citealt{DELLABRUNA20}. These would allow us to compare an independent measure of cluster reddening (via the Balmer decrement) to our SED-fitted reddening to check for consistency. Further, \citet{KAHRE18} additionally produced extinction maps for two of the LEGUS fields in our study by spatially binning the extinction of individual stars, based on isochrone matching of their broadband photometry, thus providing another check on cluster reddening. 2) We could utilize overlapping ALMA CO(2-1) maps (e.g. \citealt{LEROY21}), which would allow us to firstly determine the degree to which H$\alpha$ traces the molecular gas in a region. Secondly, we would be able to infer whether the clusters in our sample have foreground dust, which we could then use to support or oppose their SED-fitted dust reddening. 

Another key uncertainty is the use of a fixed covering fraction of 0.5 in our SED-modeling, and a careful examination could be performed to determine how the use of a proper covering fraction for each cluster (e.g Scheuermann et al., in prep.) might affect the ages we have presented here. 

Lastly, stellar populations within denser natal clouds may go undetected in H$\alpha$, thus an analysis of our sample's completeness would be critical in determining more accurate cluster populations and distributions of cluster properties, particularly for clusters with concentrated H$\alpha$ (e.g. \citealt{MESSA21}, \citealt{KIM21}). Forthcoming infrared observations of nearby galaxies with the James Webb Space Telescope, such as described by \citet{LEE21}, will provide inventories of dust embedded clusters to probe these earliest stages of star formation.


\end{enumerate}

\section*{Acknowledgements}

Based on observations with the NASA/ESA/CSA Hubble Space Telescope which were retrieved from MAST at the Space Telescope Science Institute, operated by the Association of Universities for Research in Astronomy, Incorporated, under NASA contract NAS5-26555. The observations were obtained through \textit{HST} programs $\#$13364 and $\#$13773. Support for these programs was provided through a grant from the STScI under NASA contract NAS5-26555. A.A. acknowledges the support of the Swedish Research Council, Vetenskapsr{\aa}det, and the Swedish National Space Agency (SNSA). ROD and AW acknowledge the support of UNAM via grant agreement PAPIIT no. IA-102120. ATB would like to acknowledge funding from the European Research Council (ERC) under the European Union’s Horizon 2020 research and innovation programme (grant agreement No.726384/Empire).

\section*{Data Availability}
The broadband images and star cluster catalogs used in this study are taken from LEGUS (\citealt{CALZETTI15}, \citealt{ADAMO17}), which includes WFC3 (GO-13364; F275W, F336W, F438W, F555W, and F814W) and archival ACS (F435W, F555W, F606W, and F814W) \textit{HST} imaging, and are publicly available via their website (\url{https://legus.stsci.edu}). The LEGUS-H$\alpha$ images (GO-13773; PI R. Chandar), which include the F657N narrow-band filter and F547M medium-band filter are available via MAST. The only exception is NGC 5457, whose star cluster catalogs have been obtained through private communication and are to be published in Linden et al. (in prep.); the images for NGC 5457 remain publicly available. 




\bibliographystyle{mnras}
\bibliography{main} 




\appendix
\section{}

The analyses performed throughout this work are based on the H$\alpha$ classifications of 3757 total clusters, as described in Section~\ref{sec:section3}. The complete list of classifications, including whether the cluster is isolated, is provided as supplementary online material alongside relevant identification information from the LEGUS cluster catalogs. An abbreviated version of this table is shown in Table~\ref{tab:HalphaAbbFullTable}.

\begin{table*}
\begin{tabular}{llllllll}
\multicolumn{8}{c}{\textbf{Cluster Identification Information and H$\alpha$ Classes}} \\
\hline
ObjectID & LEGUS X & LEGUS Y & RA       & Dec       & H$\alpha$ Class & Isolated? & Field   \\
\hline
\hline
3        & 3077.165 & 4717.243 & 73.57212 & -53.3508  & 3             & TRUE      & ngc1705 \\
9        & 4140     & 4315.868 & 73.55252 & -53.35522 & 3             & TRUE      & ngc1705 \\
13       & 3472.717 & 4121.713 & 73.56483 & -53.35736 & 3             & TRUE      & ngc1705 \\
15       & 3227.907 & 4082.261 & 73.56934 & -53.35779 & 3             & TRUE      & ngc1705 \\
16       & 4137.33  & 4061.561 & 73.55257 & -53.35802 & 3             & TRUE      & ngc1705 \\
26       & 3791.484 & 3955     & 73.55895 & -53.35919 & 3             & TRUE      & ngc1705 \\
42       & 3726     & 3870.536 & 73.56016 & -53.36012 & 1             & FALSE     & ngc1705 \\
53       & 3869     & 3892.138 & 73.55752 & -53.35988 & 3             & FALSE     & ngc1705 \\
65       & 4002.706 & 3777     & 73.55505 & -53.36115 & 3             & FALSE     & ngc1705 \\
72       & 4045.371 & 3869.126 & 73.55427 & -53.36013 & 3             & TRUE      & ngc1705 \\
\vdots       & \vdots & \vdots & \vdots & \vdots & \vdots             & \vdots      & \vdots \\
\hline
\hline
\end{tabular}
\caption{An abbreviated table of H$\alpha$ classifications for all 3815 clusters detected within the 21 fields in this study (1 = concentrated, 2 = partially exposed, 3 = no emission). In addition to these classes, we include whether the cluster is isolated (no neighboring clusters within 75pc), as well as relevant identification information taken from the LEGUS cluster catalogs (\url{https://legus.stsci.edu}). The full table is available online as supplementary material.}
\label{tab:HalphaAbbFullTable}
\end{table*}



\bsp	
\label{lastpage}
\end{document}